\documentclass[a4paper,11pt]{article}

\usepackage{jheppub} 
\usepackage{amsmath,amssymb,amsthm,amscd,graphicx}
\usepackage{psfrag}
\usepackage[english]{babel}
\usepackage{float}\input epsf.sty

\usepackage{color}

\theoremstyle{definition}

\newcommand{\re}{{\rm e}}
\newcommand{\ri}{{\rm i}}
\newcommand{\rd}{{\rm d}}

\newcommand{\be}{\begin{equation}}
\newcommand{\ee}{\end{equation}}
\newcommand{\ba}{\begin{aligned}}
\newcommand{\ea}{\end{aligned}}
\newcommand{\ben}{\begin{eqnarray}\displaystyle}
\newcommand{\een}{\end{eqnarray}}

\newcommand{\sectiono}[1]{\section{#1}\setcounter{equation}{0}}

\newdimen\tableauside\tableauside=1.0ex
\newdimen\tableaurule\tableaurule=0.4pt
\newdimen\tableaustep
\def\phantomhrule#1{\hbox{\vbox to0pt{\hrule height\tableaurule width#1\vss}}}
\def\phantomvrule#1{\vbox{\hbox to0pt{\vrule width\tableaurule height#1\hss}}}
\def\sqr{\vbox{%
  \phantomhrule\tableaustep
  \hbox{\phantomvrule\tableaustep\kern\tableaustep\phantomvrule\tableaustep}%
  \hbox{\vbox{\phantomhrule\tableauside}\kern-\tableaurule}}}
\def\squares#1{\hbox{\count0=#1\noindent\loop\sqr
  \advance\count0 by-1 \ifnum\count0>0\repeat}}
\def\tableau#1{\vcenter{\offinterlineskip
  \tableaustep=\tableauside\advance\tableaustep by-\tableaurule
  \kern\normallineskip\hbox
    {\kern\normallineskip\vbox
      {\gettableau#1 0 }%
     \kern\normallineskip\kern\tableaurule}%
  \kern\normallineskip\kern\tableaurule}}
\def\gettableau#1{\ifnum#1=0\let\next=\null\else
\squares{#1}\let\next=\gettableau\fi\next}

\title{\huge{Matrix models for topological strings: exact results in the planar limit}}

\author{Szabolcs Zakany}

\affiliation{D\'epartement de Physique Th\'eorique\\
Universit\'e de Gen\`eve, Gen\`eve, CH-1211 Switzerland}

\emailAdd{Szabolcs.Zakany@unige.ch} 

\abstract{We study the large $N$ expansion of a family of matrix models related to topological strings on toric Calabi-Yau threefolds. These matrix models compute spectral observables of underlying operators obtained by quantizing the mirror curves. 
They have the form of a deformed $O(2)$ matrix model, with a specific non-polynomial potential involving the Faddeev quantum dilogarithm. Their planar limit is studied using a particular conformal mapping depending on two parameters, from which several universal results can be obtained. 
As expected, the spectral curves controlling the planar limit of the matrix models are the mirror curves themselves, which in our cases have genus $1$. 
Our results encompass all those toric geometries with genus $1$ mirror where an explicit one-cut matrix integral is known: local $\mathbb P^2$, local $\mathbb F_0$, local $\mathbb F_2$, and degenerations of the resolved $\mathbb C^3/\mathbb Z_5$, the resolved $\mathbb C^3/\mathbb Z_6$ and the resolved $Y^{3,0}$ geometries amongst others.}

\begin{document}
\maketitle

\sectiono{Introduction}  

Studying aspects of string theories using matrix models has a long tradition
\cite{Douglas:1990,Brezin:1990rb,Gross:1989vs,Kontsevich:1992,Dijkgraaf:2002fc}.
Recently, a conjectural relationship linking the topological string on toric Calabi-Yau threefolds with the spectral theory of a trace class operator on the real line was stated \cite{Grassi:2014zfa} and its various aspects extensively studied 
\cite{Codesido:2015dia,Marino:2015ixa,Kashaev:2015wia,Wang:2015wdy,Sun:2016obh,Hatsuda:2015qzx,Franco:2015rnr,Marino:2016rsq,Couso:2016vwq,Grassi:2017qee,Marino:2018elo,Grassi:2018bci}. It has in particular provided a new family of matrix models which conjecturally encode the all genus topological string free-energies in an appealing way: the topological string free-energies in the conifold frame emerge in a 't Hooft-like limit of these well defined, convergent matrix models \cite{Marino:2015ixa}.
In this sense, the relationship can be viewed as a non-perturbative realization of the closed topological string on these geometries.

The relationship between topological strings and spectral theory (often called the TS/ST correspondence) proposed in \cite{Grassi:2014zfa} is a very concrete conjecture in the spirit of large $N$ dualities. A review can be found in \cite{Marino:2015nla}. 
Starting with a toric Calabi-Yau threefold\footnote{We focus on the case where the toric Calabi-Yau threefold has a genus one mirror curve.}, the target geometry of topological string theory, we define the operator $\mathsf O$ as the appropriately quantized version of its mirror curve. The mirror curve is given by the vanishing locus of $O(\re^x,\re^y)+\kappa$, where $O(X,Y)$ is a bivariate polynomial for toric geometries. 
The quantization is performed by essentially replacing the variables $x,y$ by the quantum mechanical operators $\mathsf x, \mathsf y$, satisfying the canonical commutation relation $[\mathsf x,\mathsf y]=\ri \hbar$.
The inverse operator $\rho=\mathsf O^{-1}$ is a trace class operator, whose associated spectral quantities conjecturally encode the topological string data. In particular, when considered in a 't Hooft-like limit,
\be
	\label{tHooftregime}
	N,\hbar \rightarrow \infty, \qquad \frac{N}{\hbar}=\lambda \qquad \text{fixed},
\ee
 the fermionic spectral traces $Z_N$ should encode the standard topological string genus $g$ free-energies $F_g(\lambda)$ in the conifold frame:
\be
	\label{conj}
	 \log Z_N = \sum_{g=0}^\infty \hbar^{2-2g} \mathcal F_g(\lambda).
\ee
  The quantity $\lambda$, often called the 't Hooft coupling, is then identified with a flat coordinate on moduli space which vanishes at the conifold locus. 
  
  The fermionic spectral traces $Z_N$ can often be cast in the form of convergent matrix models if an integral kernel representation $\rho(x_1,x_2)$ can be found for the operator $\rho$.
In some cases, as shown in \cite{Kashaev:2015kha}, the integral kernel $\rho(x_1,x_2)$ can indeed be obtained, and the conjectural relationship with the topological string free-energies can be explicitly checked. For these examples, this was done pertubatively in the small 't Hooft coupling limit \cite{Marino:2015ixa,Kashaev:2015wia,Codesido:2015dia,Codesido:2016ixn}.
Exact planar quantities, on the other hand, could only be obtained in those cases where the matrix model takes the form of an undeformed $O(2)$ matrix model written in terms of an integral over eigenvalues \cite{Kashaev:2015wia}.

However, in many instances, the matrix model rather takes the form of a deformed $O(2)$ model, where the eigenvalue interaction term has the same form as for matrix models associated to the six-vertex model \cite{Kostov:1999qx}, or to a certain Leigh-Strassler deformation of supersymmetric gauge theories \cite{Dijkgraaf:2002dh,Dorey:2002pq}.
An important difference with respect to those cases is that here the potential term is non polynomial and depends non trivially on the inverse string coupling $\hbar = g_s^{-1}$. 
It involves the Faddeev (or modular) quantum dilogarithm, and in the 't Hooft limit its leading term essentially reduces to standard dilogarithms.

Still, the techniques of \cite{Kostov:1999qx} can be adapted to our case. In this work, we show how to solve the exact planar limit of a family of matrix models which has a deformed $O(2)$ interaction term. This family of matrix models contains the family of four term operators of \cite{Codesido:2016ixn}, and another new family of four term operators. The first family reduces to the family of three term operators of \cite{Kashaev:2015kha}.
 The planar one-point correlation function (or more precisely a twisted version of it) is obtained using a useful conformal map between a two-cut sphere and a flat torus, as in \cite{Kostov:1999qx}, and also \cite{Hoppe:1999xg,Kazakov:1998ji}. 
 This conformal map involves two parameters which are fixed by the precise form of the potential. Then, several other planar quantities can be derived where the potential dependance only enters through these two parameters (this is very similar to the hermitian matrix model, where the dependance on the potential enters through the endpoints of the eigenvalue distributions). 
 We work out these general formulas for the second derivative of the planar free-energy and the planar connected two-point correlator. We also retrieve the special geometry relations between the period integrals on the underlying torus and the 't Hooft coupling and the derivative of the planar free-energy.
 
 The full solution of the matrix model in the t' Hooft limit would require the knowledge of all the $n$-point correlators at all orders in the 't Hooft expansion, or an algorithmic way to construct them. For many matrix models such as the hermitian matrix model or the $O(n)$ matrix model, this is given by the topological recursion \cite{Eynard:2004mh,Chekhov:2005rr,Eynard:2007kz}. The topological recursion is an algorithmic procedure, which, given a set of ``initial conditions", generates the full set of the 't Hooft expanded $n$-point correlators recursively. The ``initial conditions" are essentialy the spectral curve and the planar two point function.
 The fact that the topological string amplitudes for all toric geometries obey the topological recursion is well known as the BKMP theorem. It was conjectured in \cite{Marino:2006hs,Bouchard:2007ys}, first proved in \cite{Eynard:2012nj}, and the proof further formalized and extended in \cite{2013arXiv1310.4818F,2016arXiv160407123F}. It then appears that if we could manage to show that the matrix models considered in this paper were to satisfy the topological recursion with the right ``initial conditions", the BKMP theorem would imply the proof of the relation between our matrix models and the topological string free-energies.\footnote{We would like to thank M. Mari\~no for pointing this out.} This work can be seen as establishing the ``initial condition" part of this program. The remainding task would be to show that these matrix models satisfy the topological recursion, which may perhaps be inferred from the results of \cite{Borot:2013lpa} (where loop equations of matrix models with rather general eigenvalue interaction terms are studied and shown to be solved by the topological recursion).\footnote{Although, the definitions of the $n$-point functions entering the topological recursion in that setup are not precisely those we consider here.}
 
 
 This paper is organised as follows. 
 The matrix integral is presented in section \ref{section2}, and the saddle-point equation for the distribution of eigenvalues is derived. 
 In section \ref{planarOnePointSection}, the saddle-point equation is solved in the 't Hooft limit through the twisted planar one-point correlator $\varpi_{1,0}(X)$. The weak 't Hooft limit is considered and the emergence of the spectral curve is studied in examples. 
 In section \ref{section4}, universal results for the derivative $\partial_\lambda \varpi_{1,0}(X)$ and the planar free-energy are derived, and checked in the weak and strong 't Hooft coupling limit in several examples. The derivative of the planar free-energy $\mathcal F_0'(\lambda)$ and the 't Hooft coupling $\lambda$ given by period integrals on the underlying torus.  In section \ref{section5}, with the help of the loop insertion operator, a universal expression for the twisted planar two-point correlator $\varpi_{2,0}(X_1,X_2)$ is obtained, which is checked in an example in the weak 't Hooft coupling limit. 
 In Appendix \ref{appA}, we gather important properties of the Faddeev quantum dilogarithm, which are used in Appendix \ref{appB} to derive the matrix models from the quantized mirror curve operators. The technique for obtaining perturbative expansions in the weak 't Hooft coupling are outlined in Appendix \ref{appC}. In Appendix \ref{appD}, we derive a useful relation valid for our model for the 't Hooft coupling $\lambda$ as a function of the modular $\tau$ parameter, whereas Appendix \ref{appE} contains an expression which did not fit in the main part.

\sectiono{The matrix model and the saddle point equation}  
\label{section2}

Let $\mathsf x, \mathsf y$ be canonically conjugate operators such that $[\mathsf x, \mathsf y]=\ri \hbar$. Let also $\alpha,\beta,\gamma$ and \mbox{$m \geq 0,n >0$} be real numbers.
Our starting point is a family of operators given by
\be
	\label{operatorO}
	\mathsf O =   \re^{\mathsf x'}+\re^{\mathsf y'}+\re^{\gamma} \re^{-m \mathsf x'-n \mathsf y'} \\
	+(\re^{\alpha}+\re^{\beta}) \re^{-(m+1)\mathsf x'-(n-1)\mathsf y'}+\re^{\alpha+\beta}\re^{-2(m+1) \mathsf x'-(2n-1)\mathsf y'}.
\ee
This operator family generalizes the three-terms operators of \cite{Kashaev:2015kha}, and the four-term operators of \cite{Codesido:2016ixn}.
As shown in Appendix \ref{appB}, the inverse operator
\be
	\rho=\mathsf O^{-1}
\ee
has an integral kernel which can be obtained in an appropriate representation.
Some special cases related to genus 1 toric Calabi-Yau geometries and degenerations of higher genus geometries are highlighted in Appendix \ref{appB}.

The fermionc spectral traces $Z_N$ are defined to be the coefficients of the Fredholm determinant of $\rho$:
\be
	{\rm det}(1+\kappa   \rho ) = 1+\sum_{N=1}^\infty \kappa^N Z_N.
\ee
It is known since Fredholm how to express these coefficients in terms of the integral kernel $\rho(\nu_1,\nu_2)$:
\be
	\label{ZnDet}
\ba
	Z_N &=  \frac{1}{N!} \int_{\mathbb R^N} \rd^N \nu  \,\, \underset{i,j=1,...,N}{\rm det} \,\,  \rho(\nu_i,\nu_j) \\
\ea
\ee
Using the explicit form of $\rho(\nu_1,\nu_2)$ given in (\ref{rhoOfnu}) and the Cauchy determinant formula (\ref{CauchyDet}), this can be rewritten in a form better suited to study the 't Hooft limit.
This form is what we will call the matrix model representation of the fermionic spectral traces:
\be
\label{ZNdef}
	Z_N =  \frac{1}{N!} \int_{\mathbb R^N} \frac{\rd^N \nu}{(2\pi )^N} \re^{-\hbar \sum_{k=1}^N  {\rm Re} V(\nu_k) } \frac{\prod_{i>j}(2\sinh \frac{\nu_i-\nu_j}{2})^2}{\prod_{i,j} 2\cosh( \frac{\nu_i-\nu_j}{2}-\ri \pi C)},
\ee
where
\be
	C=\frac{m-n+1}{2(m+n+1)},
\ee
and the potential term $\rm Re V(\nu)$ is the real part of $V(\nu)$, which is given in terms of the Faddeev quantum dilogarithm in eq. (\ref{Vfull}). See Appendix \ref{appA} for its definition and some of its properties.
It is a rather unusual potential since it is non-polynomial (not even meromorphic) and it has a non-trivial $\hbar$ dependance.
The eigenvalue interaction term is given by the ratio of hyperbolic functions, which is precisely what appears in the study of the six-vertex model \cite{Kostov:1999qx} (where it is solved for a certain potential which is polynomial in $\re^{\nu_k}$). It is a deformation of the interaction term of the $O(n)$ matrix model introduced in \cite{Kostov:1988fy}, with $n=2$ (the $O(n)$ case is solved in \cite{Eynard:1995nv,Eynard:1995zv,Borot:2009ia}). The $O(2)$ case is retrieved for $C=0$. The same kind of interaction term also appears in matrix models computing the effective superpotentials of supersymmetric gauge theories with a certain Leigh-Strassler deformation, as in \cite{Dijkgraaf:2002dh,Dorey:2002pq}. In that context, the multicut case with polynomial potential was studied in \cite{Benini:2004nn}. In the context of $\mathcal N=1^*$ 5D gauge theory, the matrix integral with the deformed $O(2)$ interaction and a certain rational potential (rational in $\re^{\nu_k}$) is considered in \cite{Hollowood:2003gr}.

The convergent version of the deformed $O(2)$ matrix model for generic potential can be defined in the following way for $\beta \in (0,2\pi)$ \cite{Kostov:1999qx}:
\be
	\label{betaMM}
	\hat{ \mathcal Z}_{N,\beta} = \int \rd A \int \rd M^\dagger \rd M {\rm exp} \left [ -{\rm Tr}\left ( \ri \re^{-\ri \frac{\beta}{2} }A M M^\dagger- \ri \re^{\ri \frac{\beta}{2} }A M^\dagger M+\hbar \mathcal W(A)\right ) \right ],
\ee
where the integral over $A$ is an integral over all $N \times N$ hermitian matrices with \emph{positive} eigenvalues, with measure
$\rd A =\prod_{i=1}^N \rd {\rm Re} A_{ii} \prod_{i<j} \rd {\rm Re} A_{ij}\rd {\rm Im} A_{ij}$. The integral over $M$ is an integral over all $N \times N$ complex matrices with measure $\rd M^\dagger \rd M =\prod_{i,j=1}^N \rd {\rm Re} M_{ij}\rd {\rm Im} M_{ij}$. The parameter $\hbar$ is positive, and the function $\mathcal W(z)$ is bounded from below on the positive real line and goes to $+\infty$ when $z \rightarrow 0,\infty$. Then, integrating over the complex matrix $M$ and reducing the integration over $A$ to an integration over its eigenvalues using the usual techniques, we find that
\be
	Z_N = \frac{G(N+1)}{2^N \pi^{\frac{N(N+1)}{2}+N^2}} \hat { \mathcal Z}_{N,\beta}, \qquad \qquad \beta=2\pi \left (\frac{1}{2}-C \right ),
\ee
where $\mathcal W(\re^\nu) = {\rm Re} V(\nu)$ has the appropriate behaviour, see below. The function $G(N)$ is the Barnes $G$-function defined by $G(N+1)=\Gamma(N)G(N)$, $\Gamma(1)=1$. In all practical computations in the following, the form given in (\ref{ZNdef}) will be more convenient. 

The link with topological strings on toric Calabi-Yaus is obtained by considering (\ref{ZNdef}) in the 't Hooft-like regime
(\ref{tHooftregime}).
In this limit, the potential can be replaced by its  large $\hbar$ asymptotic expansion given in (\ref{Vexpansion}--\ref{Vkpart}).
Then, the asymptotic expansion of the free-energy $ \log Z_N $
is conjecturally given by (\ref{conj}), where $\mathcal F_g(\lambda)$ are the genus $g$ free-energies of the standard topological string (in the conifold frame) on the underlying toric Calabi-Yau threefold. In this interpretation, $\lambda$ is a flat coordinate on moduli space which vanishes at the conifold point. We therefore want to study our matrix model in the 't Hooft limit.

As usual for large $N$ matrix models, we perform a saddle point analysis of the multidimensional integral for $Z_N$. Let us write
\be
	Z_N=  \frac{1}{N!} \int_{\mathbb R^N} \rd^N \nu \, \re^{-N^2 S_{\rm eff}}.
\ee 
The effective action is given by
\be
	\label{Seff}
	S_{\rm eff} = \frac{1}{N \lambda} \sum_{k=1}^N {\rm Re}\, V(\nu_k)-\frac{1}{2 N^2} \sum_{i \neq j} \log 4 \sinh^2 \left (  \frac{\nu_i-\nu_j}{2} \right )+\frac{1}{N^2}\sum_{i,j} \log 2 \cosh \left (\frac{\nu_i-\nu_j}{2}-\ri \pi C \right ).
\ee
In the large $N$ limit, $Z_N$ is given by its value at the saddle point configuration given by \mbox{$\nu^*=(\nu_1^*,...,\nu_N^*)$} satisfying for all $k=1,...,N$:
\be
	\label{speq1}
\ba
	0 = \left. \frac{\partial S_{\rm eff}}{\partial \nu_k} \right |_{\nu=\nu^*} &= \frac{1}{N \lambda} \frac{\partial}{\partial \nu_k^*} {\rm Re}V(\nu_k^*) \\
	& \quad 
	-\frac{1}{2N^2} \sum_{i \neq k} \left [ 2\coth\left (  \frac{\nu_k^*-\nu_i^*}{2} \right ) -\tanh \left (  \frac{\nu_k^*-\nu_i^*}{2}  -\ri \pi C\right )-\tanh \left (  \frac{\nu_k^*-\nu_i^*}{2}  + \ri \pi C\right ) \right ].
\ea
\ee
It is convenient to use the exponential of $\nu_k^*$ as variables. Define
\be
	X_k = \re^{\nu_k^*}.
\ee
In the 't Hooft limit (\ref{tHooftregime}), only the leading part of the potential denoted $V_{0}(\nu)$ contributes, given by (\ref{V0part}). Let us set
\be
	\mathcal V (X_k) = {\rm Re} V_0(\nu_k^*).
\ee
The explicit expression is
\be
	\label{explicitW}
\ba
	\mathcal V(X)  &= -\frac{1}{2\pi} \log X-\frac{m+n+1}{4\pi^2 \ri} \left [ { \rm Li_2} \left (-X\re^{\tilde \alpha+\frac{\ri \pi n}{m+n+1} } \right ) - { \rm Li_2} \left (-X\re^{\tilde \alpha-\frac{\ri \pi n}{m+n+1} } \right ) \right. \\
	& \qquad  \qquad \qquad \qquad \qquad \qquad 
	+ { \rm Li_2} \left (-X\re^{\tilde \beta+\frac{\ri \pi n}{m+n+1} } \right ) - { \rm Li_2} \left (-X\re^{\tilde \beta-\frac{\ri \pi n}{m+n+1} } \right ) \\
	& \qquad  \qquad \qquad \qquad \qquad \qquad  \left.
	+ { \rm Li_2} \left (-X\re^{\tilde \gamma+\frac{\ri \pi (m+1)}{m+n+1} } \right ) - { \rm Li_2} \left (-X\re^{\tilde \gamma-\frac{\ri \pi(m+1)}{m+n+1} } \right ) \right ],
\ea
\ee
where $(\tilde \alpha,\tilde \beta, \tilde \gamma)=\frac{2\pi}{m+n+1}\hbar^{-1}(\alpha,\beta,\gamma)$.
The saddle point equation (\ref{speq1}) at leading order in large $N$ can be rewritten as
\be
	\label{speq2}
	\frac{1}{\lambda} \mathcal V'(X_k) = \frac{1}{N} \sum_{i \neq k} \left [ \frac{2}{X_k-X_i}-\frac{\omega}{\omega X_k-X_i}-\frac{\omega^{-1}}{\omega^{-1} X_k-X_i} \right ], \qquad k=1,...,N,
\ee
where 
\be
	\omega = -\re^{2\pi \ri C}.
\ee
We will always suppose that $\omega$ is different from $1$ (indeed, for  $C = \pm 1/2$ the matrix model is not well behaved in the 't Hooft limit).

\sectiono{The planar one-point function}  
\label{planarOnePointSection} 

\subsection{The planar one-point function in an elliptic paramterization}

In the 't Hooft limit (\ref{tHooftregime}), we expect the saddle point values $\nu_i^*$ to condense in a region $\mathcal I$ of the complex plane in such a way that they can be modelled by a distribution of a continuous variable. We work here in the variables $X=\re^\nu$, and use the normalized eigenvalue density $\rho(X)\rd X$, such that
\be
	\int_{\mathcal I} \rho(X) \rd X =1.
\ee
Using this distribution, the sum in (\ref{speq2}) can be rewritten as an integration on the region $\mathcal I$:
\be
	\label{speq3}
	\frac{1}{\lambda} \mathcal V'(X) =2 {\rm P} \int_{\mathcal I}  \frac{\rho(X') \rd X'}{X-X'}-  \omega \int_{\mathcal I} \frac{\rho(X') \rd X'}{\omega X-X'}-\omega^{-1} \int_{\mathcal I} \frac{\rho(X') \rd X'}{\omega^{-1} X-X'}, \qquad X \in \mathcal I,
\ee
where $\mathrm P$ denotes the Cauchy principal value. It can be checked in our cases that the potential has a single minimum, so we expect the saddle point configuration to condense inside a single interval on the positive real line 
\be
	\mathcal I = [a,b], \qquad {\rm with} \qquad a,b \in \mathbb R_+ \qquad {\rm and} \qquad a<b. 
\ee
As in \cite{Marino:2018elo}, let us define the following $n$-point correlators:
\be
	\label{nCorrW}
	W_n(X_1,...,X_n) = \Big \langle  \prod_{k=1}^n \sum_{i_k=1}^N \left ( \frac{1}{X_k-\re^{u_{i_k}}}-\frac{1}{X_k- \omega \re^{u_{i_k}}} \right ) \Big \rangle^{(c)},
\ee
where the expectation value is defined in (\ref{expVal}) and $(c)$ means the connected correlator. 
We will often find more useful the following, slightly different set of $n$-point correlators
\be
	\label{twistedCorr}
\ba
	\varpi_n(X_1,...,X_n) &= ( \omega^{1/2} X_1) \cdots ( \omega^{1/2} X_n) W_n(\omega^{1/2} X_1,...,\omega^{1/2} X_n) 
	\\
	&= \Big \langle  \prod_{k=1}^n \sum_{i_k=1}^N \left ( \frac{ \omega^{1/2}X_k}{\omega^{1/2}X_k-\re^{u_{i_k}}}-\frac{\omega^{-1/2} X_k}{\omega^{-1/2}  X_k-  \re^{u_{i_k}}} \right ) \Big \rangle^{(c)},
\ea
\ee
which have nicer reality properties and turn out to be more directly related to the spectral curve of the matrix integral. They will sometimes be called the twisted correlators.
The $n$-point correlators have the following 't Hooft expansion:
\be
\ba
	W_{n}(X_1,...,X_n) &= \sum_{g=0}^\infty \hbar^{2-2g-n}W_{n,g}(X_1,...,X_n), \\
	\varpi_n(X_1,...,X_n) &= \sum_{g=0}^\infty \hbar^{2-2g-n}\varpi_{n,g}(X_1,...,X_n).
\ea
\ee
In particular, the one-point correlators in the large $N$ limit can be also given in terms of the saddle-point configuration $\rho(X)$.
\be
	\varpi_{1,0}(X) = \lambda \int_{\mathcal I} \rd X' \rho(X') \left ( \frac{\omega^{1/2} X}{ \omega^{1/2} X-X'}- \frac{\omega^{-1/2}X}{\omega^{-1/2} X-X'}  \right ).
\ee
It is a function on the complex plane with branch cuts, which can be chosen to lie along the rotated intervals
\be
	\omega^{1/2}\mathcal I = [\omega^{1/2}a,\omega^{1/2}b]
	\qquad \text{and} \qquad
	\omega^{-1/2}\mathcal I = [\omega^{-1/2}a,\omega^{-1/2}b],
\ee
and analytic everywhere else.
The discontinuities along the branch cuts are given by the eigenvalue density $\rho(X)$: for $X \in \mathcal I$ we have
\be
\ba
	\label{rhoFrom1point}
	\varpi_{1,0}(\omega^{ \mp 1/2}(X - \ri 0)) - \varpi_{1,0}(\omega^{ \mp 1/2}(X + \ri 0)) &=  \pm2 \pi \ri \lambda X \rho(X),
\ea
\ee
from which we immediately obtain the condition
\be
	\label{normCond0}
	\lambda = \pm \frac{1}{2\pi \ri} \oint_{\omega^{\mp 1/2} \mathcal I} \varpi_{1,0}(X) \frac{\rd X}{X},
\ee
which is just the normalization of the eigenvalue density $\rho(X)$.
Also, from its definition, we have the following small and large $X$ behaviour:
\be
\ba
	\varpi_{1,0}(X) &= O(X) \qquad \quad  \text{for } X \rightarrow 0, \\
	\varpi_{1,0}(X) &= O(X^{-1}) \qquad  \text{for } X \rightarrow  \infty, \\
\ea
\ee
The saddle-point equation (\ref{speq3}) can be written in terms of $\varpi_{1,0}(X)$ instead of the distribution $\rho(X)$. We obtain the following condition on $\varpi_{1,0}(X)$:
\be
	\label{discontEq}
\ba
	\varpi_{1,0}(\omega^{-1/2}(X \pm \ri 0)) - \varpi_{1,0}(\omega^{1/2}(X \mp \ri 0)) &=X\mathcal V'(X), \qquad X\in \mathcal I, \\		
\ea
\ee

The key argument is that these conditions together with the different analytic properties of $\varpi_{1,0}(X)$ fix it completely. To actually find what it is, we essentially use the technique of \cite{Kostov:1999qx}, adapted to our potential function $X \mathcal  V'(X)$. In order for the technique of \cite{Kostov:1999qx} to work for solving (\ref{discontEq}), we need $\mathcal V'(X)$ to be a meromorphic function of $X$. In the present situation this is not the case, since $\mathcal V(X)$ is built from dilogarithm functions ${\rm Li}_2(z)$, so $X \mathcal V'(X)$ has logarithmic singularities. Therefore, what we do is we basically differentiate all involved functions another time and solve for the derivative of $\varpi_{1,0}(X)$. Indeed, the right hand side of (\ref{discontEq}) becomes a meromorphic function. Concretely, let us define $U(X)$ to be a meromorphic function satisfying
\be
	 X(X \,\mathcal V'(X))' = U(\omega^{1/2}X)-U(\omega^{-1/2}X).
\ee
(an expression for $U(X)$ is given below). We also define
\be
	\label{JandUrel}
\ba
	J(X) &= U(X) + X \varpi_{1,0}'(X)
\ea
\ee
In terms of $J(X)$, by appropriate multiplication by $X$ and differentiation, we can rewrite (\ref{discontEq}) as
\be
	\label{Jperiodic}
\ba
	J(\omega^{\pm 1/2}(X+\ri 0)) - J(\omega^{\mp 1/2}(X-\ri 0)) &= 0
\ea
\ee
valid for $X\in \mathcal I$. So we learn that $J(X)$ is a function on the complex plane with two cuts along $\omega^{1/2}\mathcal I$ and $\omega^{-1/2}\mathcal I$, with a sort of ``periodic" matching along the different sides of the two branch cuts as a consequence of (\ref{Jperiodic}). These branch points can be of the inverse square root type because of the extra differentiation. Apart from these points, $J(X)$ is a meromorphic away from the cuts, all of whose poles are inherited exclusively from the potential part $U(X)$. Also, for large $X$,
\be
	J(X) = U_{\rm pol}(X)+O(X^{-1}),
\ee
where $U_{\rm pol}(X)$ is the polynomial part of $U(X)$.
These properties of $J$ tell us that if we can find a conformal map $X(u)$ from the interior of the fundamental rectangle with side $1$ and $\tau \in \ri \mathbb R_{>0}$ to the complex plane minus the cuts, then the function
\be
	j(u) \equiv J(X(u))
\ee
is an elliptic function. In other words, it is a doubly periodic meromorphic function in the complex plane, with fundamental domain given by the rectangle with sides 1 and $\tau$. An elliptic function is completely determined by its poles and its polar behaviour at those poles, up to an additive constant (a consequence of Liouville's theorem).
Define the odd $\vartheta$-function as
\be
	\vartheta_1(u) =\frac{1}{\ri} \sum_{n \in \mathbb Z}(-1)^n \re^{\ri \pi (n+\frac{1}{2})^2\tau+2\pi \ri (n+\frac{1}{2})u},
\ee
satisfying
\be
\ba
	\vartheta_1(u+1) &= \vartheta_1(-u) = -\vartheta_1(u),  \\
	\vartheta_1(u+\tau) &=- \re^{-\ri \pi \tau-2\pi \ri u} \vartheta_1(u) \\
	\vartheta_1(u) &=\vartheta_1'(0) u +O(u^3).
\ea
\ee
\begin{figure}[t]
  \centering
    \includegraphics[width=1\textwidth]{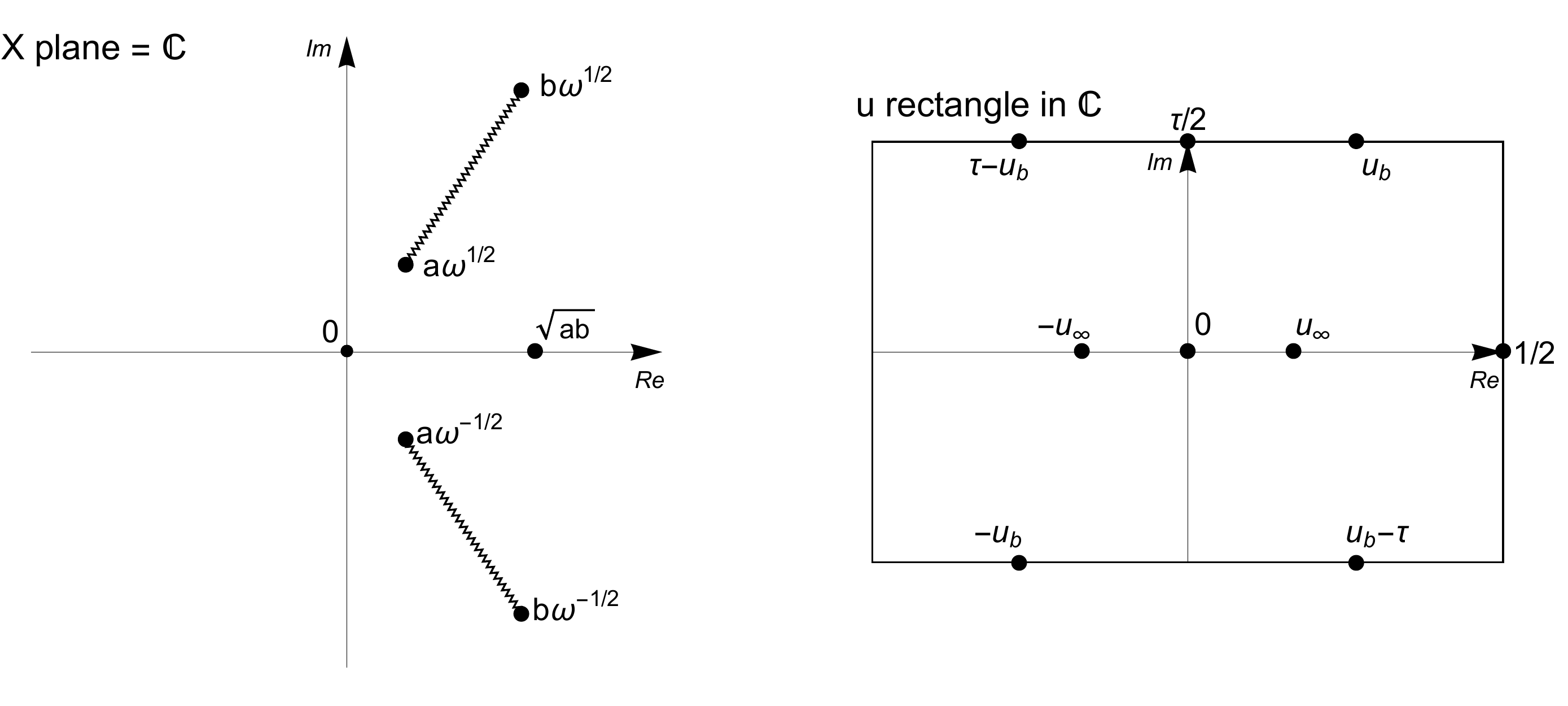} 
      \caption{The various points of interest in the $X$ plane and their images in the $u$ rectangle under the conformal mapping $X(u)$. On the left, the wiggly lines denote the segments $\omega^{1/2} \mathcal I =[\omega^{1/2}a,\omega^{1/2}b]$ and  $\omega^{-1/2} \mathcal I =[\omega^{-1/2}a,\omega^{-1/2}b]$, which are the branch-cuts of $\varpi_{1,0}(X)$. These segments are mapped to the upper and lower sides of the $u$-rectangle. This conformal mapping is a mapping between a single sheet in the cut $X$ plane to a flat torus. On this flat torus $j(u)=J(X(u))$ is a well defined meromorphic function. So $j(u)$ is an elliptic function of $u$. 
      \label{XofuFig}}
\end{figure}
An elliptic function with poles at $u=B_1,...,B_k$ can be written as a ratio of $\vartheta_1$ functions:
\be
	f(u) = {\rm const } \times  \prod_{\ell=1}^k \frac{\vartheta_1(u-A_\ell)}{\vartheta_1(u-B_\ell)},
\ee
provided that 
\be
	A_1+A_2+...+A_k = B_1+B_2+...+B_k \quad {\rm mod} \quad 1.
\ee
Actually, all elliptic functions can be written in this way.

The conformal mapping $X(u)$ has been found in \cite{Kostov:1999qx,
Hoppe:1999xg}, and can also be written using the $\vartheta_1$ function. In our notation, it is given by
\be
	X(u) = \sqrt{ab} \frac{\vartheta_1(u_\infty+u)}{\vartheta_1(u_\infty-u)}.
\ee
The values $a,b$ determine the interval $\mathcal I$ and $u_\infty$ is the point in the $u$ plane that is mapped to $\infty$. Correspondingly, the point $-u_\infty$ is mapped to $X=0$. See Figure \ref{XofuFig}.
We want to map the upper and lower sides of the fundamental rectangle centred at $u=0$ to the cuts $\omega^{ 1/2} \mathcal I$ and $\omega^{-1/2} \mathcal I$ respectively. This is fulfilled if a shift by $\tau$ in the $u$ plane is equivalent to a multiplication by $\omega$ in the $X$ plane. This fixes the relation
\be
	\label{XshiftRule}
	\omega=\frac{X(u+\tau)}{X(u)} =\frac{\vartheta_1(u_\infty+u+\tau) \vartheta_1(u_\infty-u)}{ \vartheta_1(u_\infty-u-\tau) \vartheta_1(u_\infty+u)} = \re^{-4\pi \ri u_\infty}.
\ee
The map $X(u)$ has therefore only two yet unfixed parameters: $\sqrt{ab}$ and $\tau$.
The branch-points located at $\omega^{\pm 1/2}a$ and $\omega^{\pm 1/2}b$ in the $X$ plane are stationary points of the map $X(u)$. Let us define $u_b$ to be the stationary point corresponding to $\omega^{1/2}b$:
\be
	X(u_b) = \omega^{1/2}b, \qquad  X'(u_b)=0.
\ee
The second equation is equivalent to
\be
	0 =   \frac{X'(u_b)}{X(u_b)} =\frac{\vartheta_1'}{\vartheta_1}(u_\infty+u_b)+\frac{\vartheta_1'}{\vartheta_1}(u_\infty-u_b)
\ee
So $-u_b$ is also stationary point, in fact it is the one corresponding to $\omega^{-1/2}a$.
Since $X(-u) = ab/X(u)$, the value $u_b$ defined from the above relation determines the ratio
\be
	\frac{b}{a}=\omega^{-1} \left ( \frac{\vartheta_1(u_\infty+u_b)}{\vartheta_1(u_\infty-u_b)} \right )^2,
\ee
which corresponds to the length of the cut.

To find the function $j(u)$ we will need the explicit form of the potential function $U(X)$. 
From (\ref{explicitW}) we have
\be
\ba
	X (X \mathcal V'(X))' &= \frac{m+n+1}{4\pi^2 \ri} \left (  \frac{\re^{\ri \pi \frac{n}{m+n+1} }X }{\re^{- \tilde \alpha}+\re^{\ri \pi \frac{n}{m+n+1} }X } -  \frac{\re^{-\ri \pi \frac{n}{m+n+1} }X }{\re^{- \tilde \alpha}+\re^{-\ri \pi \frac{n}{m+n+1} }X }  \right . \\
	& \qquad \qquad \qquad \qquad \left . 
	 + \frac{\re^{\ri \pi \frac{n}{m+n+1} }X }{\re^{-\tilde \beta}+\re^{\ri \pi \frac{n}{m+n+1} }X } -  \frac{\re^{-\ri \pi \frac{n}{m+n+1} }X }{\re^{- \tilde \beta}+\re^{-\ri \pi \frac{n}{m+n+1} }X } 
	  \right . \\
	  & \qquad \qquad \qquad \qquad \left . 
	 + \frac{\re^{\ri \pi \frac{m+1}{m+n+1} }X }{\re^{- \tilde \gamma}+\re^{\ri \pi \frac{m+1}{m+n+1} }X } -  \frac{\re^{-\ri \pi \frac{m+1}{m+n+1} }X }{\re^{-\tilde \gamma}+\re^{-\ri \pi \frac{m+1}{m+n+1} }X } 
	  \right  ).
\ea
\ee
Since
\be
	\omega = \re^{2 \pi \ri \frac{m+1}{m+n+1}} = \re^{-2 \pi \ri \frac{n}{m+n+1}},
\ee
a consistent choice for the branch of the square-root gives
\be
	\omega^{1/2} = -\re^{-\ri \pi \frac{n}{m+n+1}}=\re^{\ri \pi \frac{m+1}{m+n+1}}, \qquad  \omega^{-1/2} = -\re^{\ri \pi \frac{n}{m+n+1}}= \re^{-\ri \pi \frac{m+1}{m+n+1}}.
\ee
The function $U(X)$ is given by the relatively simple expression
\be
	U(X)=\frac{(m+n+1)}{4\pi^2 \ri} \left ( \frac{X}{\re^{-\tilde \alpha}-X} +\frac{X}{\re^{-\tilde \beta}-X}+\frac{X}{\re^{-\tilde \gamma}+X}  \right ).
\ee
Its three poles are inherited by the $J(X)$ function.
Let us list the points where $J(X)$ has a singular behaviour:
\be
\ba
	\text{single poles} \qquad : & \quad X=\re^{-\tilde \alpha}, \quad X=\re^{-\tilde \beta}, \quad X=-\re^{-\tilde \gamma},  \\
	\text{inverse square-roots} \qquad : & \quad X= \omega^{1/2}a, \quad X= \omega^{1/2}b, \quad X= \omega^{-1/2}a, \quad X= \omega^{-1/2}b.  \\
\ea
\ee
In the fundamental rectangle in the $u$ plane, we define the corresponding points $u_{\alpha,\beta,\gamma}$. 
\be
	\label{massPositions}
	X(u_\alpha)=\re^{-\tilde \alpha}, \qquad X(u_\beta)=\re^{-\tilde \beta}, \qquad X(u_\gamma)=-\re^{-\tilde \gamma}.
\ee
We also have the points $ u_b,-u_b,u_b-\tau,-u_b+\tau$ corresponding to the branch-points:
\be
	X(u_b) = \omega^{1/2}b, \quad X(-u_b) = \omega^{-1/2}a, \quad X(u_b-\tau) = \omega^{-1/2}b, \quad \quad X(-u_b+\tau) = \omega^{1/2}a.
\ee
The function $j(u)$ is an elliptic function which has poles at all these locations in the $u$-rectangle. As a ratio of $\vartheta_1$-functions, it can be given as
\be
	j(u) \propto  \frac{\vartheta_1(u-v_1)\vartheta_1(u-v_2)\vartheta_1(u-v_3)\vartheta_1(u-v_4)\vartheta_1(u-v_5)}{\vartheta_1(u-u_\alpha)\vartheta_1(u-u_\beta)\vartheta_1(u-u_\gamma)\vartheta_1(u-u_b)\vartheta_1(u+u_b)},
\ee
such that
\be
	v_5=u_\alpha+u_\beta+u_\gamma-v_1-v_2-v_3-v_4 \quad {\rm mod} \quad 1.
\ee
The constants $v_i$ for $i=1,...,4$ and the overall multiplicative constant are yet undetermined (we will not need their actual values).
From (\ref{JandUrel}), the planar one-point correlator $\varpi_{1,0}(X)$ satisfies
\be
	\varpi_{1,0}(X) = \int \frac{J(X)-U(X)}{X}\rd X = \left [ \int \frac{X'(u)}{X(u)} j(u)\rd u \right ]_{u=u(X)} -  \int \frac{U(X)}{X}\rd X.
\ee
The second integral gives a sum of logarithmic terms:
\be
	 \int \frac{U(X)}{X}\rd X =   \frac{m+n+1}{4\pi^2 \ri}\left ( \log(\re^{-\tilde \gamma}+X)- \log(\re^{-\tilde \alpha}-X )- \log(\re^{-\tilde \beta}-X)\right ),
\ee
 whereas evaluating the first integral needs a bit more investigation. Firstly,
we find that
\be
	\log X(u+1) = \log X(u), \qquad \qquad   \log X(u+\tau) = \log X(u)+\log \omega.
\ee
So $\frac{X'(u)}{X(u)}$ is an elliptic function with poles at $\pm u_\infty$ and zeros at $\pm u_b$. It can be therefore given by
\be
	\label{dXoverX}
	\frac{X'(u)}{X(u)} \propto  \frac{\vartheta_1(u-u_b)\vartheta_1(u+u_b)}{\vartheta_1(u-u_\infty)\vartheta_1(u+u_\infty)}.
\ee
Then, we have
\be
	\frac{X'(u)}{X(u)} j(u) \propto  \frac{\vartheta_1(u-v_1)\vartheta_1(u-v_2)\vartheta_1(u-v_3)\vartheta_1(u-v_4)\vartheta_1(u-v_5)}{\vartheta_1(u-u_\alpha)\vartheta_1(u-u_\beta)\vartheta_1(u-u_\gamma)\vartheta_1(u-u_\infty)\vartheta_1(u+u_\infty)}.
\ee
Let us now define
\be
\ba
	\xi(u) &= k_1 \log \vartheta_1(u+u_\infty)+k_2 \log \vartheta_1(u-u_\infty)+k_3 \log \vartheta_1(u-u_\alpha) \\
	& \qquad +k_4 \log \vartheta_1(u-u_\beta) +k_5 \log \vartheta_1(u-u_\gamma),
\ea
\ee
where $k_i$ are yet undetermined constants satisfying
\be
	\sum_{i=1}^5 k_i=0.
\ee
Since
\be
	\xi(u+1) = \xi(u), \qquad \xi(u+\tau) = \xi(u)+{\text{constant shift}},
\ee
the function $\xi'(u)$ is an elliptic function with single poles at $\pm u_\infty$ and $u_{\alpha,\beta,\gamma}$. Therefore, it can be written as
\be
	\xi'(u) =  \frac{\vartheta_1(u-v_1)\vartheta_1(u-v_2)\vartheta_1(u-v_3)\vartheta_1(u-v_4)\vartheta_1(u-v_5)}{\vartheta_1(u-u_\alpha)\vartheta_1(u-u_\beta)\vartheta_1(u-u_\gamma)\vartheta_1(u-u_\infty)\vartheta_1(u+u_\infty)}+{\rm const}
\ee
for the appropriate $k_i$.
From this, we obtain 
\be
	\int \frac{X'(u)}{X(u)} j(u) \rd u = \xi(u) + k_6 u +k_7.
\ee
So we find
\be
\ba
	\varpi_{1,0}(X) &= k_1 \log \vartheta_1(u(X)+u_\infty)+k_2 \log \vartheta_1(u(X)-u_\infty)+k_3 \log \vartheta_1(u(X)-u_\alpha)
	 \\
	& \qquad +k_4 \log \vartheta_1(u(X)-u_\beta) +k_5 \log \vartheta_1(u(X)-u_\gamma)+k_6 u(X)+k_7
	\\
	& \qquad 
	+\frac{m+n+1}{4\pi^2 \ri }\left ( \log(\re^{-\tilde \alpha}+ X)+\log(\re^{-\tilde \beta}+ X)-\log(\re^{-\tilde \gamma}- X) \right ) .
\ea
\ee
We now need to determine the seven constants $k_i$, $i=1,...,7$. Firstly, the function $\omega(X)$ is regular at the values $X=\re^{-\tilde \alpha},\re^{-\tilde \beta},-\re^{-\tilde \gamma}$. So the apparent logartihmic singularities at these points must cancel, which fixes 
\be
	-k_3 =-k_4=k_5 = \frac{m+n+1}{4\pi^2 \ri }.
\ee
From the explicit expression of the map $X(u)$, we get that
\be
\ba
	u(X) &= u_\infty -\frac{\sqrt{ab} \, \vartheta_1(2u_\infty)}{\vartheta_1'(0)}\frac{1}{X}+O(X^{-2}), \\
	u(X) &= -u_\infty +\frac{\vartheta_1(2u_\infty)}{\sqrt{ab} \, \vartheta_1'(0)}X+O(X^2). \\
\ea
\ee
Using this, we expand our expression for $\varpi(X)$. For $X \rightarrow 0$,
\be
\ba
	\varpi_{1,0}(X) &= k_1 \log X+\left [ k_1 \log \frac{\vartheta_1(2u_\infty)}{\sqrt{ab}} +k_2 \log \vartheta_1(-2u_\infty)-k_6 u_\infty+k_7 \right. \\
	& \qquad  \qquad \qquad \qquad
	\left. -\frac{m+n+1}{4\pi^2 \ri } \log \left (\re^{\tilde \alpha+\tilde \beta-\tilde \gamma} \frac{\vartheta_1(-u_\infty-u_\alpha)\vartheta_1(-u_\infty-u_\beta)}{\vartheta_1(-u_\infty-u_\gamma)}  \right ) \right ]+O(X).
\ea
\ee
For $X \rightarrow \infty$,
\be
\ba
	\varpi_{1,0}(X) &= \left [ -k_2+\frac{m+n+1}{4\pi^2 \ri } \right ] \log X+\left [ k_1 \log \theta_1(2u_\infty)+k_2 \log \sqrt{ab}\vartheta_1(-2u_\infty)+k_6 u_\infty+k_7 \right. \\
	& \qquad  \qquad \qquad \qquad
	\left. -\frac{m+n+1}{4\pi^2 \ri } \log \left( \frac{\vartheta_1(u_\infty-u_\alpha)\vartheta_1(u_\infty-u_\beta)}{\vartheta_1(u_\infty-u_\gamma)} \right )\right ]+O(X^{-1}).
\ea
\ee
But we know from the analytic structure of $\varpi_{1,0}(X)$ that we must have $\varpi_{1,0}(X)=O(X)$ for $X \rightarrow 0$, and $\varpi(X)=O(X^{-1})$ for $X \rightarrow \infty$. So we get four conditions for the four remaining constants $k_1,k_2,k_6,k_7$, given by the vanishing of the two expressions above at order $\log X$ and constant order. We find
\be
\ba
	k_1 &= 0, \\
	k_2 &= \frac{m+n+1}{4\pi^2 \ri }, \\
	k_6 &= 0, \\
	k_7 &= \frac{m+n+1}{4\pi^2 \ri } \log \left ( \re^{\tilde \alpha+\tilde \beta-\tilde \gamma} \frac{  \vartheta_1(u_\infty+u_\alpha) \vartheta_1(u_\infty+u_\beta) }{\vartheta_1(2u_\infty) \vartheta_1(u_\infty+u_\gamma)} \right ).
\ea
\ee
In the end, the twisted planar one-point correlator is
\be
\ba
	\varpi_{1,0}(X) &= \frac{m+n+1}{4\pi^2 \ri }\log \left ( \frac{ \vartheta_1(u_\infty+u_\alpha) \vartheta_1(u_\infty+u_\beta)}{\vartheta_1(2u_\infty) \vartheta_1(u_\infty+u_\gamma)} \frac{\vartheta_1(u(X)-u_\gamma)\vartheta_1(u(X)-u_\infty)}{\vartheta_1(u(X)-u_\alpha)\vartheta_1(u(X)-u_\beta)} \right. \\
	& \qquad \qquad \qquad \qquad \qquad \qquad
	\left. \times \frac{(\re^{\tilde \alpha}X-1)(\re^{\tilde \beta}X-1)}{(\re^{\tilde \gamma}X+1)} \right ).
\ea
\ee
We derived $\varpi_{1,0}(X)$ (and thus $W_{1,0}(X)=\lambda^{-1} \varpi_{1,0}( \omega^{-1/2} X)/X$) only using the saddle-point equation for the derivative function $J(X)$. In particular, we lost some information on the growth of the potential for $X \rightarrow \infty$. We need to consider the full condition (\ref{discontEq}).
We use that if $X$ is in the interval $\mathcal I$, we have
\be
	u(\omega^{-1/2}(X\pm \ri 0))+\tau = u(\omega^{1/2}X(X\mp \ri 0)).
\ee
From our expression for $\varpi_{1,0}(X)$, we find that
\be
	\varpi_{1,0}(\omega^{-1/2}(X\pm \ri 0))-\varpi_{1,0}(\omega^{1/2}(X\mp \ri 0)) -  X \mathcal V'(X) = \frac{m+n+1}{4\pi^2 \ri } \log \left (\re^{-2\pi \ri (u_\gamma+u_\infty-u_\alpha - u_\beta-\frac{1}{m+n+1})} \right ).
\ee
So the saddle-point equation for $\varpi_{1,0}(X)$ implies the extra relation 
\be
	\label{sumAllu}
	u_\gamma + u_\infty - u_\alpha -u_\beta = \frac{1}{m+n+1} \quad {\rm mod} \quad 1.
\ee
This should in principle define the relation between $\sqrt{ab}$ and $\tau$. At this stage, the only undetermined parameter is the imaginary side of the fundamental rectangle $\tau$. It is implicitly determined as a function of the 't Hooft coupling $\lambda$ from the normalization condition (\ref{normCond0}).
\newline

{\bf Summary.} Let us give a brief summary of the result for the planar one-point function. We have fixed parameters $\tilde \alpha, \tilde \beta, \tilde \gamma$ and $m,n$. We define $\omega=\re^{2\pi \ri \frac{m+1}{m+n+1}}=\re^{-2\pi \ri \frac{n}{m+n+1}}$. The interval where the eigenvalues (in the variable $X$) condense in the 't Hooft limit is given by $\mathcal I = [a,b]$. The endpoints can be obtained as a function of $\tau$ (the modular parameter of the $\vartheta_1$-function) through two relations. The first is
\be
	\label{bovera}
	 \frac{b}{a}= \omega^{-1} \left ( \frac{\vartheta_1(u_\infty+u_b)}{\vartheta_1(u_\infty-u_b)} \right )^2,
\ee
where 
\be
\ba
	u_\infty &= \frac{n}{2(m+n+1)}, \\
	0 &= \frac{\vartheta_1'}{\vartheta_1}(u_\infty+u_b)+\frac{\vartheta_1'}{\vartheta_1}(u_\infty-u_b).
\ea	
\ee
The second is the consistency relation which can be written as
\be
	\label{sumUcond}
	\re^{2\pi \ri (u_\gamma - u_\alpha - u_\beta)} =\re^{2\pi \ri  \frac{2-n}{2(m+n+1)}},
\ee
where $u_{\alpha,\beta,\gamma}$ are determined through
\be
\ba
	\re^{-\tilde \alpha} = \sqrt{ab} \frac{ \vartheta_1(u_\infty+u_\alpha)}{\vartheta_1(u_\infty-u_\alpha)}, \qquad
	\re^{-\tilde \beta} = \sqrt{ab} \frac{ \vartheta_1(u_\infty+u_\beta)}{\vartheta_1(u_\infty-u_\beta)}, \qquad
	-\re^{-\tilde \gamma} = \sqrt{ab} \frac{ \vartheta_1(u_\infty+u_\gamma)}{\vartheta_1(u_\infty-u_\gamma)}.
\ea
\ee
This gives an implicit relation for  $\sqrt{ab}$ in terms of $\tau$. 
The planar one point correlator $\varpi_{1,0}(X)$ is given by
\be
\ba
	\varpi_{1,0}(X) &= \frac{m+n+1}{4\pi^2 \ri }\log \left ( \frac{ \vartheta_1(u_\infty+u_\alpha) \vartheta_1(u_\infty+u_\beta)}{\vartheta_1(2u_\infty) \vartheta_1(u_\infty+u_\gamma)} \frac{\vartheta_1(u(X)-u_\gamma)\vartheta_1(u(X)-u_\infty)}{\vartheta_1(u(X)-u_\alpha)\vartheta_1(u(X)-u_\beta)} \right. \\
	& \qquad \qquad \qquad \qquad \qquad \qquad
	\left. \times \frac{(\re^{\tilde \alpha}X-1)(\re^{\tilde \beta}X-1)}{(\re^{\tilde \gamma}X+1)} \right ).
\ea
\ee
The exact eigenvalue distribution can be obtained from this expression using (\ref{rhoFrom1point}).
The relation that fixes the modular parameter $\tau$ is the contour integral
\be
	\label{intCond}
	\lambda = \frac{1}{2\pi \ri } \oint_{\omega^{-1/2} \mathcal I} \frac{\varpi_{1,0}(X)}{X} \rd X.
\ee

\subsection{The weak 't Hooft coupling limit}  

In the weak 't Hooft coupling $\lambda$ limit, the attraction of the eigenvalues towards the minimum of the potential beats the eigenvalue repulsion. So we expect the region where the eigenvalues condense (the interval $\mathcal I=[a,b]$) to collapse to a point, which should be the minimum of the potential. This means that,
\be
	\frac{b}{a} \rightarrow 1, \qquad {\rm when} \qquad \lambda \rightarrow 0.
\ee
Let us define 
\be
	q=\re^{\ri \pi \tau},
\ee
and expand all terms involving the elliptic $\vartheta_1$-functions in small $q$. There are well known formulas for this purpose.
We have the following expansion for $X(u)$:
\be
	\label{Xsmallq}
	\log X(u) =\log \left ( \sqrt{a b} \frac{\sin \pi (u_\infty+u)}{\sin \pi (u_\infty-u)} \right ) +4 \sum_{n=1}^\infty \frac{q^{2n}}{n(1-q^{2n})} \sin(2\pi n u_\infty) \sin(2\pi n u),
\ee
and more generally
\be
\ba
	\label{smallqUseful}
	\log \frac{\vartheta_1(u)}{\vartheta_1(v)} &= \log \frac{\sin \pi u}{\sin \pi v}-2 \sum_{n=1}^\infty \frac{q^{2n}}{n(1-q^{2n})}\left ( \cos 2\pi n u - \cos 2 \pi n v  \right),\\
	\frac{\vartheta_1'}{\vartheta_1}(u) &=\pi \frac{\cos \pi u}{\sin \pi u}+4\pi \sum_{n=1}^\infty \frac{q^{2n}}{1-q^{2n}} \sin 2\pi n u .
\ea
\ee
Inverting functions can be done order by order in $q$.
 We find that $u_b=\frac{1}{4}+\frac{\tau}{2}+O(q)$, and the ratio $b/a$ given in eq. (\ref{bovera}) is:
\be
	\frac{b}{a}=1+8 \sin(2\pi u_\infty)q +32 \sin(2\pi u_\infty)^2 q^2+O(q^3).
\ee
So we deduce that the small $q$ expansion corresponds to the weak coupling expansion (small $\lambda$ expansion) of the matrix model. Indeed, when $q \rightarrow 0$, we have that $a \rightarrow b$ and the interval $\mathcal I$ collapses to a point, which is given by $a=b=\sqrt{ab}$.

It is also useful to have a small $q$ expansion of the inverse of $X(u)$, which can be worked out order by order. We find that
\be
	\re^{2\ri \pi u(X)} = \re^{2\ri \pi u_\infty}\frac{X+\re^{-2\pi \ri u_\infty}\sqrt{ab}}{X+\re^{2\pi \ri u_\infty}\sqrt{ab}} 
	+ \frac{\re^{-2 i \pi u_\infty} \sqrt{ab}\left(-1+\re^{4 i \pi 
  u_\infty}\right)^3 X \left(X^2-ab\right)}{\left(X+\re^{2 i \pi u_\infty}\sqrt{ab}\right)^3 \left(\sqrt{ab}+e^{2 i \pi 
  u_\infty} X\right)}q^2+O(q^4).
\ee
This can be used together with condition (\ref{sumUcond}) to obtain a small $q$ expansion for $\sqrt{ab}$ as a function of $\tilde \alpha, \tilde \beta, \tilde \gamma$. Condition (\ref{sumUcond})  is then equivalent to
\be
	\label{implicitSqrtab}
	\re^{\frac{2 \pi \ri}{m+n+1}} = \frac{
	(1+\re^{\frac{\ri \pi n}{m+n+1}+\tilde \alpha}\sqrt{ab} )(1+\re^{\frac{\ri \pi n}{m+n+1}+\tilde \beta}\sqrt{ab} )(1+\re^{\frac{\ri \pi(m+1)}{m+n+1}+\tilde \gamma}\sqrt{ab} )
	}{ 
	(1+\re^{-\frac{\ri \pi n}{m+n+1}+\tilde \alpha}\sqrt{ab} )(1+\re^{-\frac{\ri \pi n}{m+n+1}+\tilde \beta}\sqrt{ab} )(1+\re^{-\frac{\ri \pi(m+1)}{m+n+1}+\tilde \gamma}\sqrt{ab} )
	 }+O(q^2).
\ee
This condition can be easily written in terms of the planar potential ${\rm Re} V_0(u)$.
The minimum of the planar potential in variable $X=\re^{\nu}$ is given by the condition
\be
	0 =\left. ( \partial_\nu {{\rm Re} V_0 }) \right |_{\nu=\log X} = X \mathcal V'(X).
\ee
Exponentiating it, we get
\be
\ba
	1 &= \re^{ 2\pi \ri \frac{2\pi}{m+n+1}  X \mathcal V'(X)} \\
	& = \re^{-\frac{2 \pi \ri}{m+n+1}}
	\frac{
	(1+\re^{\frac{\ri \pi n}{m+n+1}+\tilde \alpha}X )(1+\re^{\frac{\ri \pi n}{m+n+1}+\tilde \beta}X )(1+\re^{\frac{\ri \pi(m+1)}{m+n+1}+\tilde \gamma}X )
	}{ 
	(1+\re^{-\frac{\ri \pi n}{m+n+1}+\tilde \alpha}X )(1+\re^{-\frac{\ri \pi n}{m+n+1}+\tilde \beta}X)(1+\re^{-\frac{\ri \pi(m+1)}{m+n+1}+\tilde \gamma}X)
	 },
\ea
\ee
corresponding to the leading order of (\ref{implicitSqrtab}) with $X=\sqrt{ab}$.
This is as expected: in the small $q$ limit, the middle of the eigenvalue distribution at the position $\sqrt{ab}$ approaches the minimum of the planar potential.

In order to obtain $\sqrt{ab}$ as a function of $q$, we have to invert (\ref{implicitSqrtab}). This is quite tedious in the general case. Let us specialize for the sake of simplicity.
\newline

{\bf The three term operators.} In this case, we have $\tilde \alpha, \tilde \beta \rightarrow \infty$, and $\tilde \gamma=0$. 
Since $u_\alpha=u_\beta = u_\infty$, we have $u_\gamma=\frac{n+2}{2(m+n+1)}$, and the implicit equation for $\sqrt{ab}$ in terms of $q$ reduces to the explicit relation
\be
\ba
	 \sqrt{a b} &= \frac{ \vartheta_1(\frac{1}{m+n+1}) }{ \vartheta_1(\frac{m}{m+n+1})  }
	 	=\frac{\sin \frac{\pi}{m+n+1} }{ \sin \frac{\pi m}{m+n+1} } - 4\frac{ \sin \frac{\pi}{m+n+1} \sin \frac{\pi (m-1)}{m+n+1} \sin \frac{\pi  n}{m+n+1}}{ \sin \frac{\pi m}{m+n+1}}q^2+O(q^4).
\ea
\ee
The function $\varpi_{1,0}(X)$ simplifies to
\be
	\varpi_{1,0}(X) =  \frac{m+n+1}{4\pi^2 \ri}\log \left ( \frac{ \vartheta_1(2u_\infty) }{\vartheta_1(u_\infty+u_\gamma)} \frac{\vartheta_1(u(X)-u_\gamma)}{\vartheta_1(u(X)-u_\infty)} \frac{1}{(X+1)} \right ),
\ee
Then, we find
\be
\ba
	&\varpi_{1,0}(X) = \frac{8(m+n+1)}{\pi^2 \ri }  \\
	&    
	\times \frac{X \sin^2 \left (\frac{\pi}{m+n+1} \right ) \sin^2 \left (\frac{\pi m}{m+n+1} \right ) \sin^2 \left (\frac{\pi n}{m+n+1} \right )}
	{(X^2+X+1)-(X+1)\cos \left ( \frac{2\pi}{m+n+1} \right )-X(X+1)\cos \left ( \frac{2\pi m}{m+n+1} \right )+X\cos \left ( \frac{2\pi n}{m+n+1}  \right )}
	q^2 +O(q^4).
\ea
\ee
In order to obtain the small $\lambda$ expansion, we use the integral condition (\ref{intCond}). This can be worked out by taking the residue at $X= \omega^{-1/2} \frac{\sin \frac{\pi}{m+n+1} }{ \sin \frac{\pi m}{m+n+1} }$, and yields the small $q$ expansion $\lambda(q)$:
\be
	\label{LambdathreeTermFirst}
	\lambda = \frac{2(m+n+1)}{\pi^2} \sin \left ( \frac{\pi}{m+n+1} \right ) \sin \left (  \frac{\pi m}{m+n+1} \right ) \sin \left ( \frac{\pi n}{m+n+1} \right ) q^2+O(q^4)
\ee
(in the next section, we will give a more efficient way of finding $\lambda(q)$).
 Inverting it and plugging it back into the $q$ expansion, we obtain the  weak 't Hooft expansion for $\varpi_{1,0}(X)$:
\be
\ba
	\varpi_{1,0}(X) &= \frac{4}{\ri}  \frac{X \sin \left (\frac{\pi}{m+n+1} \right ) \sin \left (\frac{\pi m}{m+n+1} \right ) \sin \left (\frac{\pi n}{m+n+1} \right )}
	{(X^2+X+1)-(X+1)\cos \left ( \frac{2\pi}{m+n+1} \right )-X(X+1)\cos \left ( \frac{2\pi m}{m+n+1} \right )+X\cos \left ( \frac{2\pi n}{m+n+1}  \right )} \lambda \\
	& \quad +O(\lambda^2).
\ea
\ee 

This procedure can be pushed at higher order if we fix $m,n$. For example, for the matrix model related to the so called local $\mathbb P^2$ geometry, we have $m=n=1$. This is a symmetric case: $\sqrt{ab}=1$ so $b=a^{-1}$. We obtain:
\be
	\label{varpiP2ex}
\ba
	\varpi_{1,0}(X) &=\frac{\sqrt{3}}{\ri}  \frac{p_1(X)}{1+X+X^2} \lambda+ \frac{2\pi^2}{3\ri} \frac{p_2(X)}{(1+X+X^2)^3}\lambda^2
	+ \frac{8\pi^4}{27\sqrt{3}\ri} \frac{p_3(X)}{(1+X+X^2)^5} \lambda^3 \\
	&  +\frac{8\pi^6}{729 \ri} \frac{p_4(X)}{(1+X+X^2)^7} \lambda^4+\frac{32\pi^8}{10935 \sqrt{3}\ri} \frac{p_5(X)}{(1+X+X^2)^9} \lambda^5+\frac{64\pi^{10}}{295245 \ri} \frac{p_6(X)}{(1+X+X^2)^{11}} \lambda^6 \\
	& +O(\lambda^7),
\ea
\ee
where
\be
\ba
	p_1(X) &= X, \\
	p_2(X) &= X \left(X^4-X^3-6 X^2-X+1\right), \\
	p_3(X) &= X \left(X^8-5 X^7-35 X^6+16 X^5+100 X^4+16 X^3-35 X^2-5 X+1\right), \\
	p_4(X) &= -X \left(X^{12}+69 X^{11}+462 X^{10}-931 X^9-4698 X^8+441 X^7+7854 X^6+441X^5 \right. \\
   	& \qquad \qquad \left. -4698 X^4-931 X^3+462 X^2+69 X+1\right), \\
	p_5(X) &= X \left(7 X^{16}-19 X^{15}-423 X^{14}+12634 X^{13}+48719 X^{12}-48609
   X^{11}-253142 X^{10} \right. \\ 
   	& \qquad \qquad \left. -25441 X^9+289062 X^8-25441 X^7-253142 X^6-48609
   X^5+48719 X^4 \right. \\
   	& \qquad \qquad \left. +12634 X^3-423 X^2-19 X+7\right), \\
	p_6(X) &= -X \left(13 X^{20}+301 X^{19}+2596 X^{18}-58389 X^{17}-188187
   X^{16}+1476354 X^{15}+4870194 X^{14} \right. \\
   & \qquad \qquad
   \left. -2318808 X^{13}-15745323
   X^{12}-2744531 X^{11}+14859262 X^{10}-2744531 X^9 \right . \\
   & \qquad \qquad \left. -15745323 X^8-2318808
   X^7+4870194 X^6+1476354 X^5-188187 X^4 \right . \\
 & \qquad \qquad \left.   -58389 X^3+2596 X^2+301
   X+13\right).
\ea
\ee
This has been checked using the methods of Appendix \ref{appC}.
In this case, the relation (\ref{LambdathreeTermFirst}) between $\lambda$ and $q$ reads
\be
	\label{lambdaOfq}
	\lambda = \frac{9 \sqrt{3}}{4 \pi ^2} q^2+\frac{27 \sqrt{3} }{8 \pi ^2}q^4+\frac{27
   \sqrt{3}}{4 \pi ^2} q^6+\frac{117 \sqrt{3} }{16 \pi ^2}q^8+\frac{54
   \sqrt{3} }{5 \pi ^2}q^{10}+\frac{81 \sqrt{3}}{8 \pi
   ^2} q^{12}+O\left(q^{14}\right).
\ee

\subsection{The spectral curve}  

Usually when studying large $N$ matrix models, the planar one-point correlator is expressed through the so called spectral curve, which is often an algebraic curve in $\mathbb C ^2$.
 In our case, the elliptic parametrization used in the previous section hides the role of the spectral curve, but it can be made apparent and is expected to be algebraic when $m,n$ are rationals (which is the case when the matrix model is related to toric Calabi-Yau three-folds or their degenerations). 
 
 We start by considering the following functions of $u$:
 \be
 \ba
 	X(u) &= \sqrt{ab} \frac{ \vartheta_1(u_\infty+u)}{\vartheta_1(u_\infty-u)}, \\
	Y(u) &= \re^{-\frac{4\pi ^2 \ri}{m+n+1}\varpi_{1,0}(X(u))-\log \left[- \frac{\re^{\tilde \gamma}X(u)+1}{\left (\re^{\tilde \alpha}X(u)-1 \right ) \left (\re^{\tilde \beta}X(u)-1 \right )} \right ] }  \\
		&= -\frac{\vartheta_1(2u_\infty)\vartheta_1(u_\infty+u_\gamma)}{ \vartheta_1(u_\infty+u_\alpha) \vartheta_1(u_\infty+u_\beta) } \, \frac{ \vartheta_1(u-u_\alpha) \vartheta_1(u-u_\beta) }{\vartheta_1(u-u_\infty)\vartheta_1(u-u_\gamma)}.
 \ea
 \ee
 Both functions satisfy $f(u+1)=f(u)$, and moreover
 \be
 \ba
 	X(u+\tau) &= \re^{-4\pi \ri u_\infty} X(u) = \re^{-2\pi \ri \frac{n}{m+n+1}} X(u), \\
	Y(u+\tau) &= \re^{2\pi \ri (u_\alpha+u_\beta-u_\gamma-u_\infty)} Y(u) = \re^{-2\pi \ri \frac{1}{m+n+1}} Y(u),
\ea
 \ee
 where we used  to (\ref{XshiftRule}) and (\ref{sumAllu}). Therefore, we can build elliptic functions by multiplying powers of $X$ and $Y$, and an appropriate linear combination of these functions can be chosen such that all the poles vanish. By the usual argument, this combination is then a constant, and the corresponding relation defines the spectral curve.
  To see how it works, let us take some concrete cases.
   \newline 
   
 {\bf The local ${\mathbb P}^2$ case.}
 We fix $m=n=1$, $\tilde \alpha, \tilde \beta \rightarrow -\infty$, $\gamma=0$. In this case, $\sqrt{ab}=1$. The quantities $u_\infty$ and $u_\gamma$ take the special values
\be
	u_\infty =u_\alpha=u_\beta = \frac{1}{6}, \qquad \qquad u_\gamma = \frac{1}{2}.
\ee
We have
\be
	X = \frac{ \vartheta_1(u_\infty+u) }{ \vartheta_1(u_\infty-u) },
\ee
and
\be
	-1=X(u_\gamma).
\ee
Also, we have
\be
	X(u+\tau) \rightarrow  \re^{-\frac{2\pi \ri}{3}}  X(u), \qquad 	Y(u+\tau) \rightarrow  \re^{-\frac{2\pi \ri}{3}}  Y(u).
\ee
We conclude that the following three combinations are elliptic functions:
\be
\ba
	X(u)^3  & \propto  \frac{ \vartheta_1(u+u_\infty)^3 }{ \vartheta_1(u-u_\infty)^3 }, \\
	 \frac{1}{Y(u)^3} & \propto \frac{ \vartheta_1(u-u_\gamma)^3 }{ \vartheta_1(u-u_\infty)^3 }, \\[0.2cm]
	\frac{X(u)}{ Y(u) }& \propto \frac{\vartheta_1(u+u_\infty) \vartheta_1(u-u_\gamma) }{ \vartheta_1(u-u_\infty)^2 }. \\
\ea
\ee
The general combination 
\be
	\frac{1}{Y^3} + c_1 X^3+c_2 \frac{ X}{  Y}+c_3=0.
\ee
has poles of order up to three in the $u$ plane located at $u_\infty$. By appropriately fixing the constants $c_i$ for $i=1,2$, we can make all the poles cancel each other, which implies that this elliptic function is a constant. By fixing $c_3$, this constant can be made to vanish.
By looking at the polar behaviour at $u_\infty$, we find
\be
\ba
	c_1& =1,\\
	c_2 &=  \kappa, \\
	c_3 &= 1, \\
\ea
\ee
where
\be
	\kappa=3 \frac{\vartheta_1(2u_\infty)}{\vartheta_1'(0)} \left ( \frac{ \vartheta_1'}{\vartheta_1}(u_\infty-u_\gamma)- \frac{ \vartheta_1'}{\vartheta_1}(2u_\infty) \right )= -6 \frac{\vartheta_1'(\frac{1}{3})}{\vartheta_1'(0)}.
\ee
  We find the relation
  \be
  	\label{spectralcurveP2}
  	\frac{1}{Y^3} + X^3+\kappa \frac{ X}{  Y}+1=0.
  \ee
We call this the spectral curve, characterising the planar limit of the matrix model.
The appropriate branch of its solution $Y(X)$ yields $\varpi_{1,0}(X)$ which is the (twisted) planar one-point correlator:
\be
	\varpi_{1,0}(X) = -\frac{3}{4\pi^2 \ri} \left [ \log Y(X)+\log(-X-1) \right ].
\ee
 Setting 
  \be
  	\label{relLogXY}
  	\re^{x'} = \frac{Y}{X}, \qquad \re^{y'} = \frac{1}{X Y^2},
  \ee
 the spectral curve (\ref{spectralcurveP2}) becomes
  \be
  	\label{localP2xy}
  	\re^{ x'}+\re^{ y'}+\re^{-x'-y'}+\kappa=0,
  \ee
 which is indeed the mirror curve of the toric Calabi-Yau called local $\mathbb P_2$.

The planar one-point correlator itself is obtain from the twisted version through
\be
	W_{1,0}(X) = \frac{1}{\lambda} \frac{\varpi_{1,0} (\omega^{-1/2}X)}{X},
\ee
where in our case $\omega^{-1/2}=\re^{-\frac{2\pi \ri}{3}}$. From the spectral curve, we see that $Y(\re^{-\frac{2\pi \ri}{3}}X)=\re^{-\frac{2\pi \ri}{3}}Y(X)$ (we may need to change the branch of the solution $Y(X)$). We find
\be
	W_{1,0}(X) = -\frac{3}{4\pi^2 \ri \lambda} \frac{\log Y(X)}{X}+\frac{1}{2\lambda} \frac{\left. V_0'(\nu) \right |_{\nu=\log X}}{X},
\ee
where 
\be
	V_0(\nu) = -\frac{1}{2\pi} \nu +\frac{3}{2\pi^2 \ri}{\rm Li}_2(-\re^{\nu-\frac{2\pi \ri}{3}}).
\ee
is the planar potential (\ref{V0part}) in the local $\mathbb P^2$ case. This is precisely what was found empirically in \cite{Marino:2016rsq}.
The relation between the 't Hooft coupling $\lambda$ and the modulus of the spectral curve $\kappa$ is given by the normalization condition (\ref{intCond}), which reduces to:
\be
\ba
	\lambda =  \frac{3}{8\pi^3} \oint_{\omega^{-1/2} \mathcal I}  \frac{\rd X}{X} \log Y(X). 
\ea
\ee
This can be expressed using only the curve (\ref{localP2xy}). Let us call the cycle of integration the $\mathcal A$ cycle. We define the variables $x,y$ where $X=\re^x$, $Y=\re^{y}$.
The variables $x, y$ are a linearly related to $x',y'$ through (\ref{relLogXY}): $(x,y)=(-\frac{2}{3}x'-\frac{1}{3}y',\frac{1}{3}x'-\frac{1}{3}y')$. Our period integral in terms of $x'$ and $y'(x')$ solving (\ref{localP2xy}) becomes
\be
\ba
	\lambda &=  \frac{3}{(2\pi)^3} \oint_{\mathcal A} y(x)  \rd x\\
		&=
	 \frac{3}{(2\pi)^3} \oint_{\mathcal A'} \left (\frac{1}{3}x'-\frac{1}{3}y'(x') \right ) \left (-\frac{2}{3}-\frac{1}{3} \partial_{x'}y'(x') \right )  \rd x' \\
	 &=\frac{3}{(2\pi)^3}  \left [ -\frac{2}{9} \oint_{\mathcal A'} x' \rd x' +\frac{2}{9} \oint_{\mathcal A'} y'(x') \rd x'-\frac{1}{9} \oint_{\mathcal A'} x' \partial_{x'}y'(x')\rd x' +\frac{1}{9} \oint_{\mathcal A'} y'(x') \partial_{x'}y'(x')\rd x'\right ]
	  \\
	   &= \frac{1}{(2\pi)^3} \oint_{\mathcal A'} y'(x')  \rd x'.
\ea
\ee
where the $\mathcal A'$ cycle is the image of the $\mathcal A$ cycle under the linear map relating $(x,y)$ to $(x',y')$.
\newline 

  {\bf The local ${\mathbb F_0}$ case.}
  This is actually an $O(2)$ case. We fix $m=0$, $n=1$, $\tilde \beta \rightarrow -\infty$, $\gamma=0$. The quantities $u_\infty,u_\beta$ and $u_\gamma$ take the special values
\be
	u_\infty =u_\beta = \frac{1}{4}, \qquad \qquad u_\gamma =u_\alpha+\frac{1}{2},
\ee
while $u_\alpha$ is unfixed. In particular,
\be
\ba
	\re^{-\tilde \alpha} &= X(u_\alpha)=\sqrt{a b}\frac{ \vartheta_1( \frac{1}{4}+u_\alpha)}{  \vartheta_1( \frac{1}{4}-u_\alpha) }  \\
	-1 &= X(u_\gamma)=\sqrt{a b}\frac{ \vartheta_1( \frac{1}{4}+u_\gamma)}{  \vartheta_1( \frac{1}{4}-u_\gamma) } = -\sqrt{ab}  \frac{ \vartheta_1( \frac{1}{4}-u_\alpha)}{  \vartheta_1( \frac{1}{4}+u_\alpha) }, \\
\ea
\ee
from which we deduce
\be	
	a b = \re^{-\tilde \alpha}.
\ee
Also, we have
\be
	X(u+\tau) \rightarrow - X(u), \qquad 	Y(u+\tau) \rightarrow - Y(u).
\ee
This implies that the following combinations are elliptic functions:
\be
\ba
	X^{-2} &\propto \frac{\vartheta_1(u-u_\infty)^2}{\vartheta_1(u+u_\infty)^2}, \\
	Y^2 &\propto \frac{ \vartheta_1(u-u_\alpha)^2  }{ \vartheta_1(u-u_\gamma)^2}, \\
	Y X^{-1} &\propto \frac{ \vartheta_1(u-u_\alpha) \vartheta_1(u-u_\infty)}{\vartheta_1(u+u_\infty) \vartheta_1(u-u_\gamma)}, \\
	Y^2 X^{-2} &\propto \frac{ \vartheta_1(u-u_\alpha)^2 \vartheta_1(u-u_\infty)^2}{\vartheta_1(u+u_\infty)^2 \vartheta_1(u-u_\gamma)^2}.
\ea
\ee
The general combination 
\be
	\frac{1}{X^{2}}+c_1 Y^2 + c_2 \frac{Y}{X}+c_3 \frac{Y^2}{X^2}+c_4
\ee
has order two poles in the $u$ plane located at $-u_\infty$ and $u_\gamma$. By appropriately fixing the constants $c_i$ for $i=1,2,3$, we can make all the poles cancel each other, which implies that this elliptic function is a constant. By fixing $c_4$, this constant can be made to vanish.
By looking at the polar behaviour at $-u_\infty$ and $u_\gamma$, we find
\be
\ba
	c_1& =1,\\
	c_2 &=- \kappa, \\
	c_3 &= -1, \\
	c_4 &= -\re^{2\tilde \alpha},
\ea
\ee
where
\be
\ba
	 \kappa &\equiv \frac{2\vartheta_1(2u_\infty)}{\sqrt{ab}  \, \vartheta_1'(0)} \left ( \frac{\vartheta_1'}{\vartheta_1}(u_\gamma+u_\infty) - \frac{\vartheta_1'}{\vartheta_1}(u_\alpha+u_\infty)  \right ) \\
	 &= 
	 \re^{\tilde \alpha/2}  \frac{2\vartheta_1(\frac{1}{2})}{\ \vartheta_1'(0)} \left ( \frac{\vartheta_1'}{\vartheta_1} \left (u_\alpha+\frac{3}{4} \right ) - \frac{\vartheta_1'}{\vartheta_1} \left (u_\alpha+\frac{1}{4} \right )  \right ).
\ea
 \ee
  We find the relation
  \be
  	\label{spectralcurveF0}
  	\frac{1}{X^{2}}+ Y^2 - \kappa \frac{Y}{X}- \frac{Y^2}{X^2}-\re^{2\tilde \alpha} = 0,
  \ee
  The appropriate branch $Y(X)$ of the solution of this algebraic relation gives us directly the twisted planar one-point correlator as a function of $X$:
  \be
  	\varpi_{1,0}(X) =- \frac{1}{2\pi^2 \ri} \left [ \log Y (X) +\log \frac{1+X}{\re^{\tilde \alpha}X-1} \right ].
  \ee
  Setting 
  \be
  	\re^{x'} = \frac{Y}{X}, \qquad \re^{y'} = -\frac{1}{X Y},
  \ee
 the spectral curve (\ref{spectralcurveF0}) becomes
  \be
  	\re^{ x'}+\re^{2\tilde \alpha}\re^{- x'}+\re^{ y'}+\re^{- y'}+\kappa=0,
  \ee
 which is indeed the mirror curve of the toric Calabi-Yau called local $\mathbb F_0$, with mass parameter $m=\re^{2\tilde \alpha}$.
  \newline
  
We have restricted the discussion of the spectral curve to the local $\mathbb P^2$ and local $\mathbb F_0$  cases, but it should no doubt be generalizable to all the cases with $m,n$ rationals. We expect in general that the spectral curve of the matrix model is just the classical mirror curve in the appropriate variables.

\sectiono{Universal results for general potential $\mathcal V(X)$}
\label{section4}

\subsection{Universal formula for planar one-point correlator}

The conformal mapping $X(u)$ to a fundamental rectangle with sides $1$ and $\tau$ is very effective to obtain the planar one-point correlator.
 In particular, the mapping $X(u)$ is independent of the full potential $\mathcal W(X)$ of the deformed $O(2)$ matrix model (\ref{betaMM}), or its planar part $\mathcal V(X)$. It only depends on $\omega=-\re^{2\pi \ri C}=\re^{-4\pi \ri u_\infty} $, in other words on the eigenvalue interaction term. 
As for example for the hermitian matrix integrals or the standard $O(2)$ matrix model, it is possible to obtain a universal formula for the planar one-point correlator, in the sense that the potential does only enter in the formula through geometric parameters (for example the branch points). We will see that these parameters can be chosen to be the parameter $\tau$ of the conformal mapping and the midpoint $\sqrt{ab}$. Let us keep in mind that, in the beginning of this section, we deal with generic planar potential $\mathcal V(X)$ (and not, as before, with the specific case (\ref{explicitW})).

The important observation for obtaining our universal formula is that differentiating (\ref{discontEq}) with respect to $\lambda$,\footnote{We keep $X$ fixed, which means the map $u(X)$ varies since $\tau$ depends on $\lambda$.} we get a potential independent equation for $\partial_\lambda \varpi_{1,0}(X)$:
\be
	\partial_\lambda \varpi_{1,0}(\omega^{-1/2}(X\pm \ri 0))-\partial_\lambda \varpi_{1,0}(\omega^{1/2}(X\mp \ri 0)) =0.
\ee
This is the same kind of equation as (\ref{discontEq}) but for a vanishing potential.
So the function $\left. \partial_\lambda \varpi_{1,0}(X) \right |_{X=X(u)}$ in the $u$ plane must be an elliptic function which does not have poles except at the points congruent to $\pm u_b$ (the stationary points of the conformal map $X(u)$ corresponding to the branch points). Since $\varpi_{1,0}(X)$ must vanish at $X \rightarrow 0$ and $\infty$, our elliptic function $\left. \partial_\lambda \varpi_{1,0}(X) \right |_{X=X(u)}$ should vanish at $\pm u_\infty$. This determines almost completely our elliptic function:
\be
	\left. \partial_\lambda \varpi_{1,0}(X) \right |_{X=X(u)} \propto  \frac{\vartheta_1(u+u_\infty)\vartheta_1(u-u_\infty)}{\vartheta_1(u+u_b)\vartheta_1(u-u_b)}.
\ee
Comparing with (\ref{dXoverX}), we see that this can be written exclusively in terms of the conformal map $X(u)$ itself:
\be
	\left. \partial_\lambda \varpi_{1,0}(X) \right |_{X=X(u)} = c \frac{X(u)}{X'(u)}.
\ee
The constant $c$ can be determined in the following way. Differentiating the contour integral (\ref{intCond}) with respect to $\lambda$ (keeping the contour fixed), we get
\be
\ba
	1 &= \frac{1}{2\pi \ri} \oint_{\omega^{-1/2} \mathcal I} \frac{\rd X}{X} \partial_\lambda \varpi_{1,0}(X) \\
		&= \frac{1}{2\pi \ri}  \int_{\frac{1}{2}-\tau}^{-\frac{1}{2}-\tau} \rd u \frac{X'(u)}{X(u)} c \frac{X(u)}{X'(u)} \\
		& = -\frac{1}{2\pi \ri} c
\ea
\ee
So the constant $c$ equals $-2\pi \ri$. We conclude that the derivative of $\varpi_{1,0}$ with respect to $\lambda$ has a universal form
\be
\ba
	\label{dVarpiFromX}
	\partial_\lambda \varpi_{1,0}(X) &= \left. -2\pi \ri \frac{X(u)}{X'(u)} \right |_{u=u(X)} \\
	&=-\frac{2\pi \ri}{ \frac{\vartheta_1'}{\vartheta_1}(u_\infty+u(X))+ \frac{\vartheta_1'}{\vartheta_1}(u_\infty-u(X))}.
\ea
\ee
The quantity $u_\infty$ only depends on the interaction term through $\omega$, and not on the potential. The potential dependence only enters through the relation $q=\re^{\ri \pi \tau}$ and $\sqrt{ab}$ as functions of $\lambda$. The functional dependence on $\sqrt{ab}$ enters as a scaling of $X$.
In the small $q$ expansion, we find the general formula
\be
\ba
	\partial_\lambda \varpi_{1,0}(X) &=-\frac{\left (\re^{4\pi \ri u_\infty}-1 \right )\frac{X}{\sqrt{ab}} }{ \left (\re^{2\pi \ri u_\infty}+\frac{X}{\sqrt{ab}} \right ) \left (1+\re^{2\pi \ri u_\infty}\frac{X}{\sqrt{ab}} \right ) } \\
	& \quad -
	\frac{\re^{-2\pi \ri u_\infty}\left (\re^{4\pi \ri u_\infty}-1 \right )^3 \frac{X}{\sqrt{ab}} }
	{ \left (\re^{2\pi \ri u_\infty}+\frac{X}{\sqrt{ab}} \right )^3 \left (1+\re^{2\pi \ri u_\infty}\frac{X}{\sqrt{ab}} \right )^3}
	 \left [ \left ( \re^{4\pi \ri u_\infty}+1\right )\left ( \frac{X}{\sqrt{ab}}+\left ( \frac{X}{\sqrt{ab}} \right )^3 \right )  \right. \\
	& \qquad \qquad \qquad
	 \left.
	-\re^{2\pi \ri u_\infty} \left (1-6 \left (  \frac{X}{\sqrt{ab}} \right )^2+ \left (\frac{X}{\sqrt{ab}}\right )^4 \right ) \right ] q^2 +O(q^4).
\ea
\ee
As an example, we take $C=\frac{1}{6}$ corresponding to $\omega=\re^{\frac{4\pi \ri}{3}}$ and $u_\infty=\frac{1}{6}$, and $\sqrt{ab}=1$ (relevant for the local $\mathbb P^2$ case), we find the following small $q$ expansion:
\be
\ba
	\partial_\lambda \varpi_{1,0}(X) &= \frac{\sqrt{3}}{\ri} \frac{X}{(X^2+X+1)}+ \frac{3\sqrt{3}}{\ri} \frac{X(X^4-X^3-6X^2-X+1)}{(X^2+X+1)^3}q^2 \\
	& \qquad 
	+\frac{9 \sqrt{3}}{\ri} \frac{X(X^8-2 X^7-20 X^6+X^5+40 X^4+X^3-20 X^2-2 X+1)}{(X^2+X+1)^5}q^4+O(q^6). \\
\ea
\ee
Substituting the relation $q(\lambda)$ obtained by inverting (\ref{lambdaOfq}) and integrating with respect to $\lambda$, we retrieve expression (\ref{varpiP2ex}) for $\varpi_{1,0}(\lambda)$.

\subsection{The planar free-energy}

Using the result (\ref{dVarpiFromX}), we can derive universal formulas for the planar free energy $\mathcal F_0(\lambda)$. 
 At fixed $\hbar$, we have
 \be
 	\label{logZDiff}
 	\log Z_{N+1}- \log Z_N = \hbar \mathcal F_0'(\lambda)+\frac{1}{2} \mathcal F_0''(\lambda)+O(\hbar^{-1}),
 \ee
 which should be understood as a 't Hooft expansion.
 In the planar limit, the quantity $\log Z_N$ is given by the value of the effective action $-N^2 S_{\rm eff}$ in (\ref{Seff}), evaluated at the saddle point configuration given by $\nu^*_i$, $i=1,...,N$. We now assume that when adding an extra eigenvalue (going from $N$ to $N+1$), the saddle point positions of the $N$ first eigenvalues are not changed at leading order in large $N$. The extra eigenvalue $\nu_{N+1}$ will have its saddle point position $\nu_{N+1}^*$. Let us define
 \be
 	g(\nu) = \frac{1}{N} \sum_{i=1}^N \left [ \log 4 \sinh^2 \frac{\nu-\nu_i^*}{2}- \log 2 \cosh \left ( \frac{\nu-\nu_i^*}{2}-\ri \pi C \right )- \log 2 \cosh \left ( \frac{\nu-\nu_i^*}{2}+\ri \pi C \right ) \right ].
 \ee
 Then, in the large $N$ limit, we find from (\ref{logZDiff}):
 \be
 	 \mathcal F_0'(\lambda) = -{\rm Re}V_0(\nu_{N+1}^*)+\lambda g(\nu_{N+1}^*). 
 \ee
 The saddle point equation in the large $N$ limit implies that, as a function of $\nu_{N+1}^*$, the right hand side is constant on the interval where the eigenvalues condense. So we can put $\nu_{N+1}^*$ anywhere on the cut, for example at the larger end point $\nu_{N+1}^*=\log b$. Since we have
 \be
 	g(\nu) \rightarrow 0 \qquad \text{when} \qquad \nu \rightarrow 0,
 \ee
 we can write
 \be
 	 \mathcal F_0'(\lambda) =  -{\rm Re} V_0(\nu_{N+1}^*) + \lambda \int_{\infty}^{\log b} g'(\nu).
 \ee
 But in the large $N$ limit, $g'(\nu)$ is actually known:
 \be
 \ba
 	g'(\nu) &= \frac{1}{N} \sum_{i=1}^N \left ( \frac{2\re^\nu}{\re^\nu-\re^{\nu_i^*}}-\frac{\omega \re^\nu}{\omega \re^\nu-\re^{\nu_i^*}}-\frac{\omega^{-1} \re^\nu}{\omega^{-1} \re^\nu-\re^{\nu_i^*}} \right )\\
		&= \int_{\mathcal I} \rd X' \rho(X') \left (\frac{2X}{X-X'}-\frac{\omega X}{\omega X-X'}-\frac{\omega^{-1} X}{\omega^{-1} X-X'} \right )+o(1), \\
		&= \frac{1}{\lambda} \left ( \varpi_{1,0}(\omega^{-1/2}X)-\varpi_{1,0}(\omega^{1/2}X) \right)
\ea
 \ee
 Inserting this into the previous expression and using ${\rm Re}V_0(\nu) = \mathcal V(X)$, we obtain
 \be
 \ba
 	\mathcal F_0'(\lambda) &= - \mathcal V(b) + \int_{\infty}^b \frac{\rd X}{X} \left ( \varpi_{1,0}(\omega^{-1/2}X)-\varpi_{1,0}(\omega^{1/2}X)\right )
\ea
 \ee
 By changing variables in the integrations, we end up with the following formula for the derivative of the planar free-energy:
 \be
 	\label{F0primeRes}
 	\mathcal F_0'(\lambda) = - \mathcal V(b) + \int_{\omega^{1/2}b}^{\omega^{-1/2}b} \frac{\rd X}{X}  \varpi_{1,0}(X).
 \ee
 The path from $\omega^{1/2}b$ to $\omega^{-1/2}b$ is a part of the symplectic dual cycle to the $\mathcal A$ cycle (which is the cycle surrounding the cut $\omega^{-1/2} \mathcal I$).
\begin{figure}[b]
  \centering
    \includegraphics[width=0.5\textwidth]{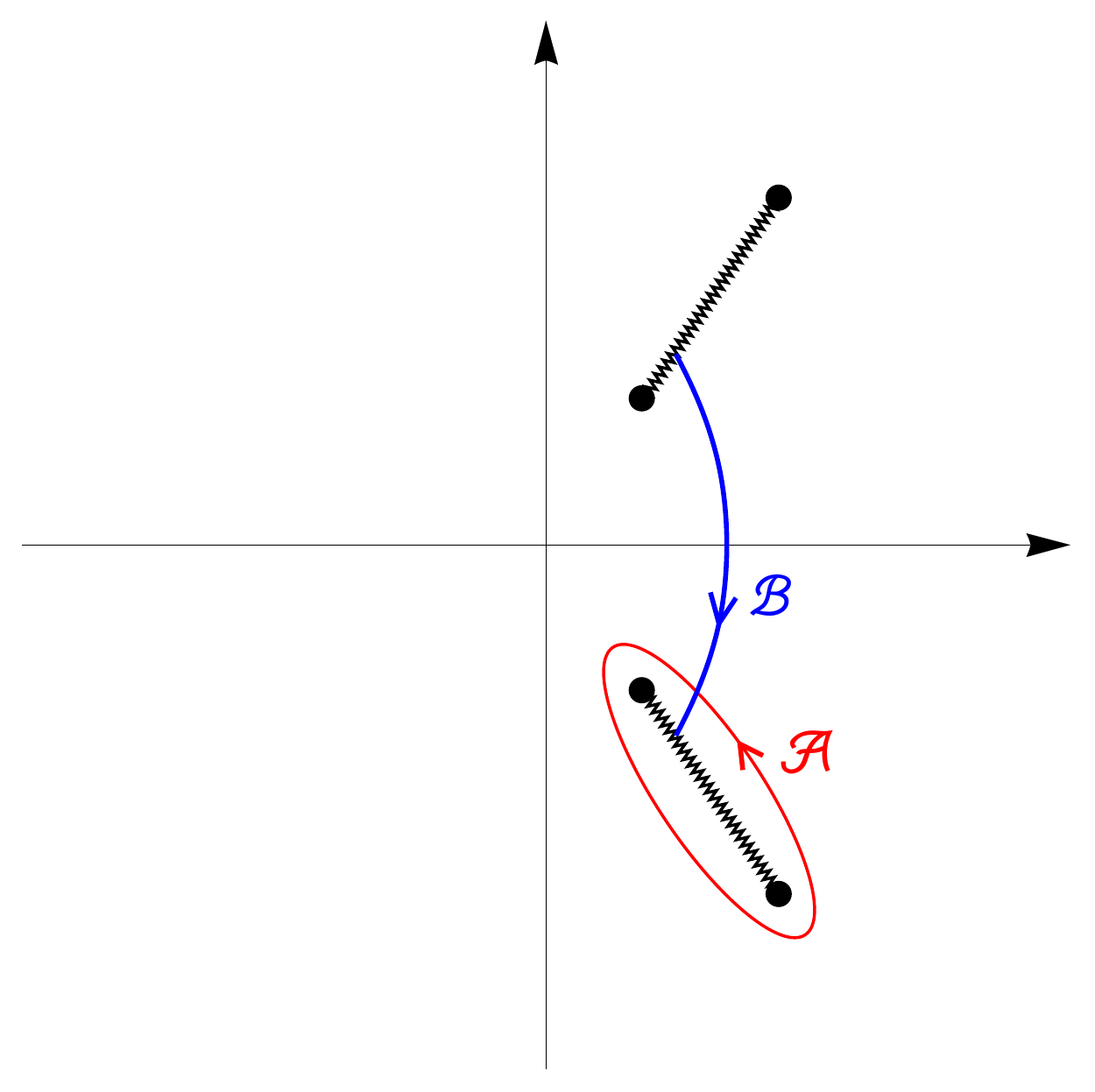} 
      \caption{The $\mathcal A$ and $\mathcal B$ cycles in the $X$ plane.
            \label{XcyclesFig}}
\end{figure}
  Let us call this contour joining the cut $\omega^{1/2} \mathcal I$ to the cut $\omega^{-1/2} \mathcal I$ the $\mathcal B$ cycle (it is not an actual cycle in the $X$ plane but it is on the torus obtained by identifying the corresponding sides of the cuts), see Fig. \ref{XcyclesFig}.  We then have the relations
 \be
 \ba
 	\lambda &= \frac{1}{2\pi \ri } \oint_{\mathcal A} \frac{\varpi_{1,0}(X)}{X} \rd X, \\
	\mathcal F_0'(\lambda) &=\int_{\mathcal B} \frac{\varpi_{1,0}(X)}{X}  \rd X - \mathcal V(b).
 \ea
 \ee
 typical of special geometry. 
As we saw in the example of local $\mathbb P^2$, the function $\varpi_{1,0}(X)$ can be substituted for $\log Y(X) $ on the spectral curve. So, as usual, the planar limit of the matrix integral is governed by the special geometry relations on the corresponding spectral curve.

 Also, from this result, we can obtain a very useful relation for the second derivative of the planar free-energy. Let us differentiate (\ref{F0primeRes}) with respect to $\lambda$. The endpoint of the integral can be chosen anywhere on the cut (not necessarily at $b$ which varies with $\lambda$), so we can keep it constant during differentiation and then set it back to the branch point $b$. We obtain
 \be
 \ba
 	\mathcal F_0''(\lambda) =  \int_{\omega^{1/2}b}^{\omega^{-1/2}b} \frac{\rd X}{X}  \partial_\lambda \varpi_{1,0}(X).
 \ea
 \ee
 We use the universal result for $\partial_\lambda \varpi_{1,0}(X)$ in  (\ref{dVarpiFromX}) and go to the $u$ plane to perform the integration:
 \be
 \ba
 	\mathcal F_0''(\lambda) &=  \int_{u_b}^{u_b-\tau}\rd u \frac{X'(u)}{X(u)}  \partial_\lambda \varpi_{1,0}(X(u)) \\
				&=  \int_{u_b}^{u_b-\tau}\rd u \frac{X'(u)}{X(u)}  \left ( \frac{-2\pi \ri X(u)}{X'(u)} \right ) \\
				&=   - 2 \pi \ri \int_{u_b}^{u_b-\tau}\rd u.
 \ea
 \ee
 We obtain a formula which is explicitly potential independent (the dependance enters implicitly through the relation $\tau(\lambda)$), and which is valid for all the deformed $O$(2) cases:
 \be
 	\label{ddF0}
 	\mathcal F_0''(\lambda) = 2\pi \ri  \tau.
 \ee
 The second derivative of the free-energy with respect to the 't Hooft coupling is the modular parameter of our elliptic parametrization. This kind of result is standard in other families of matrix models.
 
 We can find another formula involving the second derivative of the planar free-energy and the $\mathcal A$ cycle instead of the $\mathcal B$ cycle. From the matrix model, it is easy to see that the derivative of the free-energy with respect to $\hbar$ keeping $N$ fixed is
 \be
 	\partial_\hbar \log Z_{N} = - {\Big \langle}  \sum_{k=1}^N {\rm Re} V(u_k) {\Big \rangle}.
 \ee
 In the large $N$ limit, we can use the continuous distribution of eigenvalues $\rho(X)$ and the fact that $\partial_\hbar = -N^{-1}\lambda^2 \partial_\lambda$. The above relation becomes
 \be
 \ba
 	\lambda^2 \partial_\lambda  \lambda^{-2} \mathcal F_0(\lambda)&= \int_{\mathcal I} \rd X \rho(X) \mathcal V(X)  \\
	&= \frac{1}{2\pi \ri \lambda} \oint_{\omega^{-1/2}\mathcal I} \frac{\rd X}{X} \varpi_{1,0}(X) \mathcal V(\omega^{1/2}X).
\ea
 \ee
 Multiplying it by $\lambda$ and differentiating, we can use (\ref{dVarpiFromX}) to write
 \be
  \ba
 	\partial_\lambda \lambda^3 \partial_\lambda  \lambda^{-2} \mathcal F_0(\lambda)= \lambda \mathcal F_0''(\lambda) -  \mathcal F_0'(\lambda)  &= \frac{1}{2\pi \ri} \oint_{\omega^{-1/2}\mathcal I} \frac{\rd X}{X} \partial _\lambda \varpi_{1,0}(X) \mathcal V(\omega^{1/2}X), \\
	&=\frac{1}{2\pi \ri} \int_{-\frac{1}{2}-\frac{\tau}{2}}^{\frac{1}{2}-\frac{\tau}{2}} \mathcal V \left ( \omega^{1/2}X \left (u\right ) \right ) \rd u. 
\ea
 \ee
Let us differentiate the left hand side of this formula with respect to $\tau$:
\be
\ba
	\frac{\rd}{\rd \tau} \left (\lambda \mathcal F_0''(\lambda) -  \mathcal F_0'(\lambda)   \right )
	& =\left ( \mathcal F_0''(\lambda)+\mathcal F_0'''(\lambda) -  \mathcal F_0''(\lambda)   \right ) \lambda'(\tau) \\
	&= \lambda \frac{\mathcal F_0'''(\lambda)}{\frac{\rd}{\rd \lambda}\tau} \\
	&= 2\pi \ri \lambda.
\ea
\ee
In the last line, we used the universal result (\ref{ddF0}). So we obtain another universal result, this time for $\lambda$:
\be
	\label{lambdaUniversal}
	\lambda = \frac{1}{2\pi \ri} \frac{\rd}{\rd \tau} \int_{-\frac{1}{2}-\frac{\tau}{2}}^{\frac{1}{2}-\frac{\tau}{2}} \mathcal V \left ( \omega^{1/2}X \left (u\right ) \right ) \rd u.
\ee
In the above, not only the endpoints of the integral are functions of $\tau$, but also the map $X(u)$ (through $q$ and $\sqrt{ab}$ which is implicitly $q$ dependant). If $\sqrt{ab}$ is known (for example if the potential obeys $\mathcal V(X^{-1})=\mathcal V(X)$ we have $\sqrt{ab}=1$), then equations (\ref{ddF0}) and (\ref{lambdaUniversal}) determine $\mathcal F_0''(\lambda)$ completely, and also $\mathcal F_0(\lambda)$ up to two integration constants.
 
\subsection{Examples in the weak 't Hooft coupling limit}

Let us go back to our particular case, where the planar potential $\mathcal V(X)$ is given by (\ref{explicitW}).
Equation (\ref{ddF0}) gives a quite easy way of computing the planar free-energy of our models (up to the two integration constants) in the small 't Hooft coupling limit, provided we now the relation $\tau(\lambda)$. Let us consider the condition (\ref{intCond}) in the small $q=\re^{\ri \pi \tau}=\re^{\frac{1}{2} \mathcal F_0''}$ expansion. Extracting the part of $\varpi_{1,0}(X)$ which contributes non trivially to the contour integral, and then going to the $u$ variable, we find
 \be
 	\label{lambdaGeneral}
 \ba
 	\lambda &= \frac{1}{2\pi \ri} \left (\frac{m+n+1}{4\pi^2 \ri} \right) \int_{\frac{1}{2}-\frac{\tau}{2}}^{-\frac{1}{2}-\frac{\tau}{2}} \rd u \left ( \frac{\vartheta_1'}{\vartheta_1}(u_\infty+u)+\frac{\vartheta_1'}{\vartheta_1}(u_\infty-u) \right ) \left ( \log \frac{\vartheta_1(u-u_\infty)}{\vartheta_1(2u_\infty)} \right . \\
	& \qquad \qquad  \qquad \qquad \qquad
	\left.  +\log \frac{\vartheta_1(u-u_\gamma)}{\vartheta_1(u_\infty+u_\gamma)} - \log \frac{\vartheta_1(u-u_\alpha)}{\vartheta_1(u_\infty+u_\alpha)} - \log \frac{\vartheta_1(u-u_\beta)}{\vartheta_1(u_\infty+u_\beta)} \right ).
 \ea
 \ee
 \newline
 As shown in the Appendix \ref{appD}, this can be rewritten as
  \be
 	\label{lambdaNice2}
 	\lambda = \frac{m+n+1}{4\pi^2 \ri} \left (  \int_{u_\infty}^{u_\gamma}-  \int_{u_\infty}^{u_\alpha}-  \int_{u_\infty}^{u_\beta} \right )\rd u  \frac{\partial}{\partial \tau} \log  X(u).
 \ee
 In the small $q$ regime, we can use the all-order expansion (\ref{Xsmallq}) to perform the integral order by order. The outcome is
 \be
 	\lambda = \frac{m+n+1}{\pi^2} \sum_{k=1}^\infty \frac{q^{2k}}{k(1-q^{2k})^2} \sin(2\pi k u_\infty) \left ( \cos(2\pi k u_\alpha) + \cos(2\pi k u_\beta) -\cos(2\pi k u_\gamma)-\cos(2\pi k u_\infty)   \right ).
 \ee
 We need to express $u_{\alpha,\beta,\gamma}$ in terms of $\tilde \alpha,\tilde \beta$ and $\tilde \gamma$ since this relation may also be $q$-dependent. 
  Now that we have the expansion $\lambda(q)=\lambda(\re^{\frac{1}{2}\mathcal F_0''})$, we only need to invert the series and integrate twice to obtain the planar free-energy in the small t' Hooft coupling expansion. In general, we find that the expansion for $\lambda(q)$ looks like
\be
	\lambda = \sum_{k=1}^\infty { a_k q^{2k}},
\ee
from which we obtain
\be
\ba
	\mathcal F_0(\lambda) &= c_0+c_1 \lambda +\left [\frac{1}{2}\log \left(\frac{\lambda}{a_1} \right )-\frac{3}{4} \right ]\lambda^2
	-\frac{a_2}{6a_1^2}\lambda^3+\left (\frac{a_2^2}{8a_1^4} -\frac{a_3}{12 a_1^3} \right )\lambda^4 \\
	& \quad 
	+ \left (-\frac{a_2^3}{6a_1^6} + \frac{a_2 a_3}{5 a_1^5} -\frac{a_4}{20 a_1^4} \right )\lambda^5
	+ \left (\frac{7 a_2^4}{24a_1^8} - \frac{a_2^2 a_3}{2 a_1^7} + \frac{a_3^2}{12 a_1^6}+\frac{a_2 a_4}{6 a_1^6}-\frac{a_5}{30 a_1^5} \right )\lambda^6+O(\lambda^7).
\ea
\ee
The integration constants $c_0$ and $c_1$ can be obtained from the expansion of the matrix model around the minimum of the potential ${\rm Re}V_0(\nu)$. We obtain
\be
	c_0=0, \qquad c_1= - \underset{\nu \in \mathbb R}{\rm min} \, {\rm Re}V_0(\nu).
\ee
Let us look at some examples.
\newline

{\bf The three term operators.} In this case, we have $\tilde \alpha, \tilde \beta \rightarrow -\infty$, and $\tilde \gamma=0$. Therefore, $u_\alpha=u_\beta = u_\infty=\frac{n}{2(m+n+1)}$, and we have $u_\gamma=\frac{n+2}{2(m+n+1)}$. Since none of the $u_{\alpha,\beta,\gamma}$ are $q$-dependent, the $a_k$ can be obtained at all orders:
\be
	a_k = 2\frac{m+n+1}{\pi^2} \sum_{d,\ell \in \mathbb N | d\ell=k} \frac{\ell}{d} (-1)^{d-1} 
	\sin  \frac{\pi d}{m+n+1} \sin \frac{\pi d m}{m+n+1} \sin \frac{\pi d n}{m+n+1}.
\ee
The symmetry under the exchange of $m$ and $n$ is apparent.
 Also,
 \be
 	 c_1 = \frac{1}{2\pi}\log \frac{\sin \frac{\pi}{m+n+1}}{\sin \frac{m\pi}{m+n+1}} + \frac{m+n+1}{2\pi^2 } {\rm Im} \,{\rm Li}_2 \left (  \re^{-\frac{\ri \pi n}{m+n+1}  } \frac{\sin \frac{\pi}{m+n+1}}{\sin \frac{m\pi}{m+n+1}} \right ).
 \ee
 This is compatible with the results in  \cite{Marino:2015ixa}, and in particular reproduces the genus 0 free-energy in the conifold frame for local $\mathbb P^2$ ($m=n=1$).
 \newline
 
 {\bf The local $\mathbb F_0$ case.} This is actually a standard $O(2)$ case, but our formulas work perfectly well. We have $m=0, n=1,\tilde \beta \rightarrow -\infty$, and $\tilde \gamma=0$. The mass parameter of local $\mathbb F_0$ is related to $\tilde \alpha$ as $m_{\mathbb F_0}=\re^{\tilde \alpha}$. We have, $ u_\infty=u_\beta=1/4$,  and $u_\gamma=1/2+u_\alpha$. The relation between $u_\alpha$ and $\tilde \alpha$ is obtained in the small $q$ expansion using that
 \be
 	-\re^{-\tilde \alpha} = \frac{X(u_\alpha)}{X(u_\gamma)} =-\frac{\vartheta_1(1/4+u_\alpha)^2}{\vartheta_1(1/4-u_\alpha)^2},
 \ee
 and inverting in small $q$:
 \be
 	\re^{2\pi \ri u_\alpha} = \frac{1-\ri \re^{\tilde \alpha/2}}{-\ri+\re^{\tilde \alpha/2}}+\frac{8 \re^{\tilde \alpha/2}(\re^{\tilde \alpha/2}-1)}{(-\ri+\re^{\tilde \alpha/2})^3(\ri+\re^{\tilde \alpha/2})}q^2+O(q^4)
 \ee
 We obtain
\be
\ba
	a_1 &= \frac{4}{\pi^2 \cosh \left ( \frac{\tilde \alpha}{2} \right )}, \\
	a_2 &=\frac{4(3\cosh(\tilde \alpha)-1)}{\pi^2 \cosh \left ( \frac{\tilde \alpha}{2} \right )^3} , \\
	a_3 &=\frac{4(9\cosh(2\tilde \alpha)-44 \cosh(\tilde \alpha)+43)}{3\pi^2 \cosh \left ( \frac{\tilde \alpha}{2} \right )^5} , \\
	a_4 &=\frac{19\cosh(3\tilde \alpha)-310 \cosh(2\tilde \alpha)+1277\cosh(\tilde \alpha)-954}{2\pi^2 \cosh \left ( \frac{\tilde \alpha}{2} \right )^7} , \\
	...
\ea
\ee
 Also,
 \be
 	 c_1 = -\frac{\tilde \alpha}{4\pi}-\frac{1}{\pi^2} \left ( {\rm Im} \, {\rm Li}_2(\ri \re^{\tilde \alpha/2} ) + {\rm Im} \, {\rm Li}_2(\ri \re^{-\tilde \alpha/2} )\right ) = -\frac{2}{\pi^2} {\rm Im} {\rm Li}_2(\ri \re^{\tilde \alpha/2}).
 \ee
 This is compatible with the results in \cite{Kashaev:2015wia}.
 \newline
 
 {\bf The local $\mathbb F_2$ case.} This is also a standard $O(2)$ case. We have $m=0,n=1,\tilde \beta = -\tilde \alpha$, and $\tilde \gamma \rightarrow -\infty$. The mass parameter of local $\mathbb F_2$ is related to $\tilde \alpha$ as $m_{\mathbb F_2}=2\cosh(\tilde \alpha)$. 
 We find $u_\beta = -u_\alpha$ and $u_\infty=u_\gamma=1/4$. The relation between $u_\alpha$ and $\alpha$ is obtained in the small $q$ expansion using that
 \be
 	\re^{-2\tilde \alpha} =\frac{X(u_\alpha)}{X(u_\beta)}=\frac{\vartheta_1(1/4+u_\alpha)^2}{\vartheta_1(1/4-u_\alpha)^2},
 \ee
 and inverting in small $q$:
 \be
 	\re^{2\pi \ri u_\alpha} = \frac{1-\ri \re^{\tilde \alpha}}{-\ri+\re^{\tilde \alpha}}+\frac{8 (\re^{\tilde \alpha}-\re^{-\tilde \alpha})}{(-\ri+\re^{\tilde \alpha})^3(\ri+\re^{\tilde \alpha})}q^2+O(q^4)
 \ee
 Then, we obtain that the coefficients $a_i$ are given by the expressions for local $\mathbb F_0$ where we substitute $\tilde \alpha \rightarrow 2 \tilde \alpha$ (or equivalently by relating the mass parameters through $m_{\mathbb F_2}=m_{\mathbb F_0}^{1/2}+m_{\mathbb F_0}^{-1/2}$). The constant $c_1$ is given by
 \be
 	 c_1 = -\frac{1}{\pi^2} \left ( {\rm Im} \, {\rm Li}_2(\ri \re^{\tilde \alpha} ) + {\rm Im} \, {\rm Li}_2(\ri \re^{-\tilde \alpha} )\right ).
 \ee
 This is compatible with the results in \cite{Kashaev:2015wia}. Indeed, the local $\mathbb F_2$ operator is a unitary conjugate of the rescaled local $\mathbb F_0$ operator, with the right dictionary between the mass parameters \cite{Kashaev:2015wia}.
 \newline 
 
 {\bf The degeneration of $Y^{3,0}$.}
 In this case, we have $m=n=1,\tilde \beta \rightarrow -\infty$, and $\tilde \gamma=0$. We have, $ u_\infty=u_\beta=1/6$,  and $u_\gamma=1/3+u_\alpha$. The relation between $u_\alpha$ and $\tilde \alpha$ is obtained in the small $q$ expansion using that
 \be
 	-\re^{-\tilde \alpha} = \frac{X(u_\alpha)}{X(u_\gamma)} =-\frac{\vartheta_1(1/6+u_\alpha)^2}{\vartheta_1(1/6-u_\alpha)\vartheta_1(1/2+u_\alpha)},
 \ee
 and inverting in small $q$:
 \be
 	\re^{2\pi \ri u_\alpha} = \frac{(-1)^{1/6} \left(\sqrt{12 \re^{\tilde \alpha }+3}-2 \ri \re^{\tilde \alpha
   }+\ri \right)}{2 \left(\re^{\tilde \alpha }+1\right)}
   +\frac{3 (-1)^{1/6} \re^{\tilde \alpha } \left(2 \re^{\tilde \alpha }-1\right)
   \left(\sqrt{12 \re^{\tilde \alpha }+3}+3 \ri \right)^2}{4 \left(\re^{\tilde \alpha
   }+1\right)^3 \sqrt{12 \re^{\tilde \alpha }+3}}q^2+
   O(q^4).
 \ee
 \newline
  We obtain
\be
\ba
	a_1 &= \frac{9\sqrt{3}(1+4\re^{\tilde \alpha})^{1/2}}{4\pi^2(1+\re^{\tilde \alpha})} , \\
	a_2 &= \frac{27\sqrt{3} (1+9\re^{\tilde \alpha}+6 \re^{2\tilde \alpha}+16 \re^{3\tilde \alpha})}{8\pi^2(1+\re^{\tilde \alpha})^3(1+4\re^{\tilde \alpha})^{1/2}}  , \\
	a_3 &=\frac{27\sqrt{3} (1+18 \re^{\tilde \alpha}+57 \re^{2\tilde \alpha} +104 \re^{3\tilde \alpha}+216 \re^{4\tilde \alpha}-222 \re^{5\tilde \alpha}+112 \re^{6\tilde \alpha})}{4\pi^2(1+\re^{\tilde \alpha})^5(1+4\re^{\tilde \alpha})^{3/2}} , \\
	a_4 &=\frac{9\sqrt{3}}{16\pi^2(1+\re^{\tilde \alpha})^7(1+4\re^{\tilde \alpha})^{5/2}} \left ( 13+477 \re^{\tilde \alpha}+2241 \re^{2\tilde \alpha} +7419 \re^{3\tilde \alpha} +29088\re^{4\tilde \alpha}+28350 \re^{5\tilde \alpha}
	\right. \\
	& \qquad \qquad  \qquad \qquad  \qquad \qquad 
	\left. +16500 \re^{6\tilde \alpha}+250632 \re^{7\tilde \alpha}-138240 \re^{8\tilde \alpha}+16384 \re^{9\tilde \alpha} \right ),
	\\
	&...
\ea
\ee
Also,
 \be
 	 c_1 = -\frac{1}{2\pi} \log \left (m_\alpha \right )-\frac{3}{2\pi^2} \left [ {\rm Im} \, {\rm Li}_2 \left ( \frac{1-m_\alpha}{ \re^{\frac{\pi \ri}{3}} }\right ) +  {\rm Im} {\rm Li}_2 \left ( \frac{\re^{\frac{\ri \pi }{3}}}{m_\alpha} \right ) \right ],
 \ee
 where
 \be
 	m_\alpha = \frac{1+\sqrt{1+4\re^{\tilde \alpha}}}{2}
 \ee
 When $\tilde \alpha=0$, we have $(a_1,a_2,a_3,a_4)= \left (\frac{9\sqrt{15}}{8\pi^2},\,\frac{27}{2\pi^2} \sqrt{\frac{3}{5}},\, \frac{3861}{320\pi^2}\sqrt{\frac{3}{5} },\, \frac{14967}{400 \pi^2}\sqrt{\frac{3}{5} } \right )$. These results are compatible with those in \cite{Codesido:2016ixn}.
\newline

\subsection{Example in the strong 't Hooft coupling limit}
 
 The interesting feature of our exact treatment of the planar matrix model is that we can investigate the strong 't Hooft coupling regime: the large $\lambda$ limit. This regime is unavailable with the pedestrian techniques of Appendix \ref{appC}, but it is an interesting regime geometrically. Indeed, the free-energies of the standard topological string in the large-radius frame expanded around the large-radius point are generating functions for the Gromov-Witten invariants for the initial toric Calabi-Yau geometry (the geometry whose mirror curve was quantized in the first place). In terms of the coordinate $\lambda$, the large radius point is expected to be located at $\lambda \rightarrow \infty$.

We will use the $S$-dual expansions of the elliptic functions. Let us define 
\be
\tau_D=-\tau^{-1}, \qquad \qquad q_D=\re^{\ri \pi \tau_D},
\ee
and consider the small $q_D$ expansion. We use the $S$-transformation formula for the elliptic theta function:
 \be
 	\vartheta_1(u,\tau) = \frac{\ri}{\sqrt{-\ri \tau}} \re^{-\frac{\ri \pi u^2}{\tau}} \vartheta_1 \left (\frac{u}{\tau},-\frac{1}{\tau} \right ).
 \ee
 Then, we can expand (\ref{lambdaNice2}) in small $q_D$ and integrate term by term. The result is given by 
 \be
 \ba
 	\label{lambdaStrong}
 	\lambda &= \frac{m+n+1}{4\pi^2 \ri} \left ( L_D(u_\gamma)+ L_D(u_\infty)-L_D(u_\alpha)-L_D(u_\beta)  \right ),
\ea
 \ee
 where
 \be
 \ba
 	L_D(u) &=2\pi \ri \tau_D u_\infty u(u-1)+\tau_D (u+u_\infty) \log (1-\re^{2\pi \ri \tau_D (u+u_\infty) } )-\tau_D (u-u_\infty )\log(1-\re^{2\pi \ri \tau_D (u-u_\infty) }) \\
	&\qquad 
	-\frac{\ri}{2\pi }\left(   {\rm Li}_2(\re^{2\pi \ri \tau_D (u+u_\infty)} )- {\rm Li}_2(\re^{2\pi \ri \tau_D (u-u_\infty)})  \right ) \\
	&\qquad 
	 -4\tau_D \sum_{k=1}^\infty \frac{q_D^{2k}}{k(1-q_D^{2k})} (u_\infty \cos(2\pi k \tau_D u_\infty) \cos(2\pi k \tau_D u)-u \sin(2\pi k \tau_D u_\infty) \sin(2\pi k \tau_D u)) \\
	 &\qquad 
	 -4 \ri \tau_D \sum_{k=1}^\infty \frac{q_D^{2k}}{k(1-q_D^{2k})^2} \sin(2\pi k \tau_D u_\infty) \cos(2\pi k \tau_D u)\\
	  &\qquad 
	 +\frac{2}{\pi} \sum_{k=1}^\infty \frac{q_D^{2k}}{k^2(1-q_D^{2k})} \sin(2\pi k \tau_D u_\infty) \cos(2\pi k \tau_D u).
\ea
 \ee
 Since the absolute values of the real parts of $u_{\infty,\alpha,\beta,\gamma}$ are smaller or equal to $1/2$, the infinite sums are consistent small  $q_D$ expansions (series of $q_D^r$ with $r>0$).
The relation (\ref{lambdaStrong}) should be inverted using $\tau_D=-\frac{2\pi \ri}{\mathcal F_0''(0)}$. Let us look at an example. 
\newline

{\bf The local ${\mathbb P}^2$ case.} We have $m=n=1$, $\tilde \alpha, \tilde \beta \rightarrow -\infty$, and $\tilde \gamma=0$. Also, $u_\alpha=u_\beta = u_\infty=\frac{1}{6}$ and $u_\gamma=\frac{1}{2}$ are $q_D$ independent. We find
\be
	\lambda = -\frac{\tau_D^2}{36\pi}-\frac{1}{16\pi}+\frac{9-6\pi \ri \tau_D}{8\pi^3}q_D^{2/3}+\frac{9(-3+4\pi \ri \tau_D)}{8\pi^3}q_D^{4/3}+\frac{1-2\pi \ri \tau_D}{8\pi^3}q_D^{2}+O(q_D^{8/3}).
\ee
Inverting it and integrating twice, we obtain the strong 't Hooft coupling expansion of the planar free-energy in the form of a trans-series:
\be
\ba
	\mathcal F_0(\lambda) &= -\frac{4}{9}\sqrt{\pi} \hat \lambda^{3/2} +c_1 \hat \lambda +c_0-\frac{3}{16\pi^4} \re^{-4\pi^{3/2}\sqrt{\hat \lambda}}
	-\frac{9}{128\pi^5}\left (5\pi+3\pi^{-1/2} \hat \lambda^{-1/2}  \right )\re^{-8\pi^{3/2} \sqrt{\hat \lambda} } \\
	& \qquad
	-\frac{1}{384\pi^7}\left (\frac{1952 }{3}\pi^3 +\frac{1215}{2}\pi^{3/2} \hat \lambda^{-1/2}+\frac{729}{4}\hat \lambda^{-1} +\frac{243 }{16}\pi^{-3/2}\hat \lambda^{-3/2} \right )\re^{-12\pi^{3/2} \sqrt{\hat \lambda} } +...,
\ea
\ee
where
\be
	\hat \lambda = \lambda +\frac{1}{16\pi}.
\ee
The constants $c_{0,1}$ have to be found by other means. We assume that $c_1=0$.
We can also obtain the genus $0$ free-energy $F_0(t)$ in the large radius frame. We need to use the following symplectic transformation \cite{Marino:2015ixa}:
\be
	\begin{pmatrix}
	t \\ \partial_t F_0(t)
	\end{pmatrix}
	=
	\begin{pmatrix}
	-6\pi \mathcal F_0'(\lambda) \\ \frac{8\pi^3}{3}\lambda
	\end{pmatrix},
\ee
from which we deduce the function $\lambda(t)$ (by performing the trans-series inversion), as well as
\be
	\partial_t^2 F_0(t)= -\frac{4\pi^2}{9} \frac{1}{\mathcal F_0''(\lambda)}
\ee
Integrating up twice, we find
\be
	F_0(t) = \frac{t^3}{18}-3 \re^{-t}-\frac{45}{8}\re^{-2t}-\frac{244}{9}\re^{-3t}-\frac{12333}{64}\re^{-4t}+O(\re^{-5t}),
\ee
which is indeed the generating series of the genus $0$ Gromov-Witten invariants of the local $\mathbb P^2$ Calabi-Yau threefold.

  
\sectiono{Loop-insertion operator and planar connected two-point correlator}
\label{section5}

\subsection{The loop-insertion operator}
The loop insertion operator in matrix models is a very useful tool to obtain higher $n$-point correlators from lower ones, for example when considering hermitian matrix models \cite{Ambjorn:1992gw} or $O(n)$ matrix models \cite{Eynard:1995nv}.

For this section, it is convenient to write our integral in terms of $X_i=\re^{\nu_i}$:
\be
	\label{ZNintX}
	Z_{N} = \frac{\ri^{N^2}}{N!}  \int_{\mathbb R^N_+} \frac{\rd^N X}{(2\pi)^N} \re^{ -\hbar \sum_{i=1}^N \mathcal V (X_i)} \frac{ \prod_{i<j} \left (X_i-X_j \right )^2 }{\prod_{i,j} \left (\omega^{1/2}X_i-\omega^{-1/2}X_j \right ) }.
\ee
The potential $\mathcal V(X)$ has an expansion around its minimum, and we call the expansion coefficients $v_k$:
\be
	\mathcal V(X)  = \sum_{k=0}^\infty v_k X^k.
\ee
The loop-insertion operator is a differential operator, which, on quantities depending on the potential, is defined as
\be
	\frac{\delta}{\delta \mathcal V(Y)} F[\mathcal V] = \lim_{\epsilon \rightarrow 0} \frac{\left ( \left . F[\mathcal V] \right |_{v_k \rightarrow v_k-\epsilon Y^k} \right ) - F[\mathcal V]}{\epsilon}.
\ee
We promptly find that
\be
	\frac{\delta}{\delta \mathcal V(Y)} \mathcal V(X) = -\frac{Y}{Y-X}\,, 
	\qquad \qquad
	 \frac{\delta}{\delta \mathcal V(Y)} \mathcal V'(X) = -\frac{Y}{(Y-X)^2}.
\ee
We also define the twisted loop-insertion operator
\be
	\frac{\delta}{\delta_\omega \mathcal V(Y)}  = \frac{\delta}{\delta \mathcal V(\omega^{1/2}Y)}-\frac{\delta}{\delta \mathcal V(\omega^{-1/2}Y)}.
\ee
Let us apply this operator on $\log Z_N$. Using (\ref{ZNintX}), we find straightforwardly
\be
	\frac{\delta}{\delta_\omega \mathcal V(X)} \log Z_N =  \hbar \varpi_1(X),
\ee
and similarly,
\be
	\frac{\delta}{\delta_\omega \mathcal V(X_1)} \cdots \frac{\delta}{\delta_\omega \mathcal V(X_n)} \log Z_N =  \hbar^n \varpi_n(X_1,...,X_n).
\ee

\subsection{The planar connected two-point correlator}
The planar part of the connected two-point correlator $\varpi_2(X_1,X_2)$ is $\varpi_{2,0}(X_1,X_2)$. It is a symmetric function. We expect that, since it is the large $N$ limit of the two point function, it can be represented in terms of a joint connected eigenvalue probability density function $\rho_c(X_1,X_2)$ as
\be
	\varpi_{2,0}(X_1,X_2) =\lambda^2  \int_{\mathcal I} \rd X_1' \int_{\mathcal I} \rd X_2' \, \rho_c(X_1',X_2') \prod_{i=1}^2 \left ( \frac{\omega^{1/2} X_i}{ \omega^{1/2} X_i-X_i'}- \frac{\omega^{-1/2}X_i}{\omega^{-1/2} X_i-X_i'}  \right ).
\ee
Therefore, as function of one of the variables with the other fixed away from the cuts, it has the same kind of branch-cuts along $\omega^{\pm 1/2} \mathcal I$ as the planar one-point function. It does not have other singularities.

Assuming that we can commute the loop-insertion with the 't Hooft expansion, we can use the previously derived property of the loop-insertion operator to write
\be
	\frac{\delta}{\delta_\omega \mathcal V(X_2)}  \varpi_{1,0}(X_1) = \varpi_{2,0}(X_1,X_2). 
\ee
  Applying this on relation (\ref{discontEq}), we obtain the following condition on the planar two-point correlator $\varpi_{2,0}(X_1,X_2)$ for $X_1 \in \mathcal I$:
  \be
  	\varpi_{2,0}(\omega^{-1/2}( X_1\pm \ri 0),X_2)-\varpi_{2,0}(\omega^{1/2}( X_1\mp \ri 0),X_2) = -\left ( \frac{\omega^{1/2}X_1 X_2}{(\omega^{1/2}X_2-X_1)^2} - \frac{\omega^{-1/2}X_1 X_2}{(\omega^{-1/2}X_2-X_1)^2} \right ).
  \ee
  By symmetry in $X_1 \leftrightarrow X_2$, the same discontinuity equation is true for the second variable $X_2$. This is the same kind of equation as (\ref{discontEq}) with a parameter dependant meromorphic potential given by the right-hand side. So the same techniques as in section \ref{planarOnePointSection} apply. The symmetry $X_1 \leftrightarrow X_2$ further constrains the solution. Let us define
  \be
  	B(X_1,X_2) =  \varpi_{2,0}(X_1,X_2) + \frac{X_1 X_2}{(X_1-X_2)^2}.
  \ee
  It satisfies for $X_1 \in \mathcal I$
  \be
  	B(\omega^{-1/2}( X_1\pm \ri 0),X_2)-B(\omega^{1/2}( X_1\mp \ri 0),X_2) = 0
  \ee
  and similarly for $X_2$. This means that we can use again the parametrization $X(u)$ of section \ref{planarOnePointSection}. The function
  \be
  	b(u_1,u_2) = B(X(u_1),X(u_2))
  \ee
  is then an elliptic function (of periods $1$ and $\tau$) in both variables $u_1$ and $u_2$, symmetric, with the only pole located at $u_1=u_2$, which is a double pole. Expressed in terms of the $X$ variables, this double pole is of the form $\frac{X_1 X_2}{(X_1-X_2)^2}$. Such a function is well known in the study of Riemann surfaces, and it is related to the so called \emph{fundamental differential of the second kind}, sometimes also called the \emph{Bergmann kernel}. All these constraints reduce the form of $B(X_1,X_2)$ to
  \be
  	B(X_1,X_2) =  X_1 X_2 \partial_{X_1} \partial_{X_2} \log \vartheta_1(u(X_1)-u(X_2))+c,
  \ee
  where $c$ is a constant.
  The polar behaviour is assured since
  \be
  \ba
  	 \log \vartheta_1(u(X_1)-u(X_2)) &= \log \left[  {\rm const} \cdot (X_1-X_2)+O(X_1-X_2)^2 \right ] \\
	 &= \log(X_1-X_2)+O(X_1-X_2)^0,
\ea
  \ee
  and so 
  \be
  	X_1 X_2 \partial_{X_1} \partial_{X_2} \log \vartheta_1(u(X_1)-u(X_2)) = \frac{X_1 X_2}{(X_1-X_2)^2}+O(X_1-X_2)^0.
  \ee
  In the $u_i$ variables, double periodicity is guaranteed using the following rewriting:
  \be
  	b(u_1,u_2) = c+\frac{X(u_1) X(u_2)}{X'(u_1)X'(u_2)}\partial_{u_1} \partial_{u_2} \log \vartheta_1(u_1-u_2).
  \ee
 Indeed, since for $M,N \in \mathbb Z$ we have
  \be
  \ba
  	 \vartheta_1(u_1-u_2+M+N \tau) = \re^{(M+N) \pi \ri-2\pi \ri N (u_1-u_2)-N^2 \ri \pi \tau} \vartheta_1(u_1-u_2),
 \ea
  \ee
  the quantity $\partial_{u_1} \partial_{u_2} \log \vartheta_1(u_1-u_2)$
  is an elliptic function for both $u_1$ and $u_2$. Moreover, we already saw that $\frac{X'(u)}{X(u)}$
  is also an elliptic function so the full expression for $b(u_1,u_2)$ is elliptic in both variable. The constant $c$ is fixed to $0$ by sending one of the variables $X_1$ or $X_2$ to $\infty$ and requiring the expression for $B(X_1,X_2)$ to vanish. We finally find the universal formula for the planar two-point connected correlator
  \be
  \ba
  	\varpi_{2,0}(X_1,X_2)  &= X_1 X_2   \partial_{X_1} \partial_{X_2} \log \vartheta_1(u(X_1)-u(X_2))- \frac{X_1 X_2}{(X_1-X_2)^2} \\
		&=  (X_1 \partial_{X_1})(X_2 \partial_{X_2}) \log \frac{\vartheta_1(u(X_1)-u(X_2))}{X_1-X_2}.
\ea
  \ee
 It is universal in the same sense as $\partial_\lambda \varpi_{1,0}(X)$ is: the dependance on the potential only enters through $\tau$ and $\sqrt{ab}$ via the elliptic theta function and the parametrization $u(X)$.
 
 We can work out the small $q$ expansion of $\varpi_{2,0}(X_1,X_2)$. We find:
 \be
 	\label{finalvarpi20}
 \ba
 	\varpi_{2,0}(X_1,X_2) &=-\frac{(\re^{4\pi \ri u_\infty}-1)^4 \frac{X_1}{\sqrt{ab}}\frac{X_2}{\sqrt{ab}} \left (\frac{X_1}{\sqrt{ab}}-1 \right )^2\left (\frac{X_2}{\sqrt{ab}}-1 \right )^2 }
	{\prod_{i=1}^2 \left ( \re^{2\pi \ri u_\infty }+ \frac{X_i}{\sqrt{ab}}  \right )^2 \left (1+ \re^{2\pi \ri u_\infty } \frac{X_i}{\sqrt{ab}}  \right )^2} q^2 \\
	& \quad
	-\frac{(\re^{4\pi \ri u_\infty}-1)^4 \re^{8\pi \ri u_\infty}
	\sum_{i,j=1}^7 \left ( \frac{X_1}{\sqrt{ab}} \right )^i  \left ( \frac{X_2}{\sqrt{ab}} \right )^j P_{ij}(\cos 2\pi u_\infty)}
	{\prod_{i=1}^2 \left ( \re^{2\pi \ri u_\infty }+ \frac{X_i}{\sqrt{ab}}  \right )^4 \left (1+ \re^{2\pi \ri u_\infty } \frac{X_i}{\sqrt{ab}}  \right )^4} q^4 +O(q^6).
 \ea
 \ee
 The functions $P_{ij}(x)$ in the above expression are polynomials and can be found in Appendix \ref{appE}.
 
 In the case of the local $\mathbb P^2$ matrix model, we have $u_\infty=\frac{1}{6}$, $\sqrt{ab}=1$ and $q$ is given in terms of $\lambda$ by the inverse of (\ref{lambdaOfq}). We then find
 \be
 \ba
 	\varpi_{2,0}(X_1,X_2) &= -\frac{4 \pi ^2   X_1 X_2 \left({X_1}^2-1\right) \left({X_2}^2-1\right)}{\sqrt{3}
   \left({X_1}^2+{X_1}+1\right)^2 \left({X_2}^2+X_2+1\right)^2}\lambda \\
   		& \quad 
		-\frac{8\pi^4  \sum_{i,j=1}^7 X_1^i X_2^j M_{ij}}{9 \left({X_1}^2+{X_1}+1\right)^4 \left({X_2}^2+X_2+1\right)^4 }\lambda^2+O(\lambda^3),
\ea
 \ee
 where the matrix $M$ whose entries are $M_{ij}$ is given by
 \be
 	M=\left(
\begin{array}{ccccccc}
 5 & 2 & -22 & -8 & 10 & -2 & -3 \\
 2 & -8 & -44 & 0 & 44 & 8 & -2 \\
 -22 & -44 & -44 & 48 & 116 & 44 & 10 \\
 -8 & 0 & 48 & 64 & 48 & 0 & -8 \\
 10 & 44 & 116 & 48 & -44 & -44 & -22 \\
 -2 & 8 & 44 & 0 & -44 & -8 & 2 \\
 -3 & -2 & 10 & -8 & -22 & 2 & 5 \\
\end{array}
\right).
 \ee
 This (as well as the next order in $\lambda$) has been checked against the gaussian expansion of the matrix model using the techniques of Appendix \ref{appC}.
 
\sectiono{Conclusion}  
We considered a family of matrix models in the 't Hooft limit, relevant in the context of the topological string on toric Calabi-Yau threefolds. The exact solution for the planar free-energy, the planar one and two point correlators were obtained using a suitable conformal mapping between a two cut sphere and a flat torus. The planar free-energies of these matrix integrals reproduce indeed the closed topological string free-energies of the corresponding geometries in the conifold frame. 
Therefore, the genus 0 Gromov-Witten invariants of the toric Calabi-Yau threefolds can also be extracted from the strong coupling limit of the matrix models.

Exact results for higher genus free-energies and $n$-point correlators should also be studied, and it should be possible to obtain them, for example through loop equations as in other matrix models. Eventually, it would be very interesting to show that the free-energies and the $n$-point correlators obey the topological recursion of Chekov-Eynard-Orentin \cite{Eynard:2007kz}, since it is known that the topological free-energies and $n$-point amplitudes do obey them \cite{Bouchard:2007ys, Eynard:2012nj}, although in a priori different coordinates than what is considered here.

Also, we restricted the discussion to one-cut matrix integrals corresponding to toric Calabi-Yau threefolds with genus $1$ mirror curves. Multi-cut matrix integrals are also known for some higher genus case \cite{Codesido:2015dia,Codesido:2016ixn,Bonelli:2017ptp}, and an exact planar solution for them is also desirable.

 \section*{Acknowledgements}
The author would like to thank Marcos Mari\~no for many helpful comments and giving the initial impulse for this project, as well as Alba Grassi and Rinat Kashaev for useful discussions and correspondence. 

This work is is supported in part by the Fonds National Suisse, 
subsidies 200021-156995 and 200020-141329, and by the NCCR 51NF40-141869 ``The Mathematics of Physics'' (SwissMAP).

\appendix
\sectiono{Some properties of the Faddeev quantum dilogarithm}  
\label{appA}

In this Appendix, we gather some formulas concerning the Faddeev quantum dilogarithm.
The Faddeev quantum dilogarithm is an important function in the context of quantized mirror curves. It closely related to the double sine, and the double gamma function.

Let $b$ be a complex number such that ${\rm Re}(b)>0$.
The Faddeev quantum dilogarithm can be defined through the integral formula
\be
	\Phi_b(u)  = {\rm exp} \left ( \int_{{\mathbb R}+\ri 0} \frac{\re^{-2\ri y u}}{4\sinh(by)\sinh(y/b)} \frac{\rd y}{y} \right ).
\ee
The integration path is along the real axis and avoids the pole at $y=0$ by going above it. This integral definition converges in the strip
\be
	|{\rm Im}(u)| < {\rm Re}\left (\frac{b+b^{-1}}{2} \right ),
\ee
but it can be analytically continued to a meromorphic function on the complex plane. It has manifestly the following property:
\be
	\label{Phiduality}
	\Phi_b(u) = \Phi_{b^{-1}}(u). 
\ee
Also, for real or unitary $b$, we have
\be
	\overline{\Phi_b(u)} = \frac{1}{ {\Phi_b(\bar u)}}. 
\ee
By deforming the path of integration and picking up the residue at $y=0$, it is easy to obtain the following formula:
\be
	\Phi_b(-u)=\re^{\ri \pi u^2+\frac{\ri \pi}{12} \left (b^2+b^{-2}\right )} \frac{1}{\Phi_b(u)}.
\ee
Its asymptotic behaviour is \cite{Andersen:2011bt}
\be
\ba
	\Phi_b(u) & \sim 1 ,  \qquad \qquad \qquad \qquad\,\, \text{when  } \quad {\rm Re}(u) \ll 0, \\
	\Phi_b(u)  & \sim \re^{\ri \pi u^2+\frac{\ri \pi}{12} \left (b^2+b^{-2}\right )} , \qquad  \text{when  } \quad {\rm Re}(u) \gg 0.
\ea
\ee
It obeys the following difference equation
\be
	\label{diffPsi1}
	\frac{\Phi_b \left  (u-\frac{\ri b}{2  } \right)}{\Phi_b \left (u+\frac{\ri b}{2} \right )} = 1+\re^{2\pi b u},
\ee
which can be obtained from the definition, using that
\be
	\int_{\mathbb R+\ri 0} \frac{ \re^{-2 \ri u y} }{2 \sinh(y/b)} \frac{\rd y}{y} = -\log (1+\re^{2\pi b u}).
\ee
Similarly, using property (\ref{Phiduality}), we have
\be
	\label{diffPsi2}
	\frac{\Phi_b \left (u-\frac{\ri}{2b} \right )}{\Phi_b \left (u+\frac{\ri}{2b} \right )} = 1+\re^{ \frac{2\pi u}{b} }.
\ee
The poles and the zeros of the Faddeev quantum dilogarithm $\Phi_b(u)$ are located at
\be
\ba
	{\rm poles: } & \qquad u = \ri \frac{b+b^{-1}}{2} + m \ri b + n \ri b^{-1}, \\
	{\rm zeros: } & \qquad u = -\ri \frac{b+b^{-1}}{2} - m \ri b - n \ri b^{-1}, \\
\ea
\ee
for $m,n \in \mathbb Z_{\geq0}$.

The following small $b$ asymptotic expansion is also useful \cite{Andersen:2011bt}:
\be
	\log \Phi_b \left (\frac{x}{2\pi b} \right ) \sim \frac{1}{2\pi \ri} \sum_{k=0}^\infty (-4\pi^2)^k b^{4k-2} \frac{(2^{-2k+1}-1) B_{2k}}{(2k)!} {\rm Li}_{2-2k}(-\re^x),
\ee
where $B_{k}$ are the Bernoulli numbers and ${\rm Li}_k(z)$ are the polylogarithms. Using property (\ref{Phiduality}), there is the corresponding large $b$ expansion:
\be
	\label{largebAsymp}
	\log \Phi_b \left (\frac{b x}{2\pi} \right ) \sim \frac{1}{2\pi \ri} \sum_{k=0}^\infty (-4\pi^2)^k b^{-4k+2} \frac{(2^{-2k+1}-1) B_{2k}}{(2k)!} {\rm Li}_{2-2k}(-\re^x).
\ee

\sectiono{Deriving the matrix models using the Faddeev quantum dilogarithm}  
\label{appB}

In this Appendix, we show how to factorize and invert some quantized mirror curves. We use the methods of \cite{Kashaev:2015kha}  which are further used in \cite{Kashaev:2015wia} and \cite{Codesido:2016ixn}.
Let us introduce canonically commuting hermitian operators $\mathsf x$ and $\mathsf y$ such that
\be
	[\mathsf x, \mathsf y] = \ri \cdot  2\pi b^2.
\ee
with $b$ real. Suppose that $A, B$ are real numbers and $\epsilon<0$. For any function $f(z)$ which is analytic on the strip  $ -2\pi b^2 A-\epsilon < {\rm Im}(z) < \epsilon$ if $A$ is positive, or $-\epsilon < {\rm Im}(z) < -2\pi b^2  A+\epsilon$ if $A$ is negative, we have
\be
	\label{shiftx}
	\re^{A \mathsf y} f(\mathsf x) \re^{-A \mathsf y} = f(\mathsf x - 2\pi \ri b^2 A),
\ee
which follows from Taylor expanding $f(\mathsf x)$.
Similarly, for any function $g(z)$ which is analytic on the strip $-\epsilon < {\rm Im}(z) < 2\pi b^2  B+\epsilon$ if $B$ is positive, or $ 2\pi b^2  B-\epsilon < {\rm Im}(z) < \epsilon$ if $B$ is negative, we have
\be
	\label{shifty}
	\re^{B \mathsf x} g(\mathsf y) \re^{-B \mathsf x} = g(\mathsf y+2\pi \ri b^2 B).
\ee
Using these formulas, we can write the operator versions of the difference equations obeyed by the Faddeev quantum dilogarithm. 
From (\ref{diffPsi1}), we obtain
\be
\ba
	1+\re^{\mathsf x} &= \re^{\mathsf y/2} \Phi_b \left ( \frac{1}{2\pi b}\mathsf x\right ) \re^{-\mathsf y/2} \re^{-\mathsf y/2} \frac{1}{\Phi_b \left ( \frac{1}{2\pi b}\mathsf x\right )}  \re^{\mathsf y/2} \\
				&=  \re^{-\mathsf y/2} \frac{1}{\Phi_b \left ( \frac{1}{2\pi b}\mathsf x\right )} \re^{\mathsf y/2} \re^{\mathsf y/2} \Phi_b \left ( \frac{1}{2\pi b}\mathsf x\right )  \re^{-\mathsf y/2}.
\ea
\ee
From this, we obtain the formula for the conjugation of the exponentials by the unitary operator given by the dilogarithm:
\be
\ba
	\label{conjExpx}
	  \Phi_b \left ( \frac{1}{2\pi b}\mathsf x\right ) \re^{-\mathsf y} \frac{1}{ \Phi_b \left ( \frac{1}{2\pi b}\mathsf x \right )} &=  \re^{-\mathsf y}+\re^{\mathsf x- \mathsf y}, \\
	   \frac{1}{\Phi_b \left ( \frac{1}{2\pi b}\mathsf x\right )} \re^{\mathsf y}  \Phi_b \left ( \frac{1}{2\pi b}\mathsf x \right ) &=  \re^{\mathsf y}+\re^{\mathsf x+ \mathsf y}.
\ea
\ee
Repeated application of these formulas on the operator $1+\re^{\mathsf y}$ yields the basic operator relations which we then map to the desired quantum curve using a linear canonical transformation. This provides a factorization of the quantum curve, in which form it is straightforward to invert. 
We remark that in this procedure the explicit use of the pentagon identity for the Faddeev quantum dilogarithm (as in for example \cite{Kashaev:2015wia}) is not needed. 
We also remark that this procedure in itself cannot yield all the genus one toric quantum curves (let alone higher genus quantum curves). 
Indeed, it can be seen that the only curves we can reach are those whose Newton polygon (in variables $\re^{\mathsf x}$ and $\re^{\mathsf y}$) can be mapped by a linear canonical transformation to a trapezoid (a 4-gon with two parallel sides), or its degenerate case, the triangle. 
In particular, the quantum curve of local $\mathbb F_1$ has four sides with no parallel pairs of sides, so it cannot be mapped to a trapezoid using a linear canonical transformation; this is an example we cannot factorize with the present method, and should be, a priori, studied as a degenerate case of the higher genus construction of \cite{Codesido:2015dia}.

The cases which interest us can be regrouped into the single formula obtain by three successive applications of the formulas (\ref{conjExpx}). This provides a generalization of all the cases studied in \cite{Kashaev:2015kha,Kashaev:2015wia,Codesido:2016ixn}. Let us define
\be
\ba
	F(\mathsf x) &= \re^{-\frac{1}{2(m+n+1)} \mathsf x} \frac{ \Phi_b \left (\frac{\mathsf x+\alpha}{2\pi b}-\frac{n}{2(m+n+1)}\ri b \right )  \Phi_b \left (\frac{\mathsf x+\beta}{2\pi b}-\frac{n}{2(m+n+1)}\ri b \right  ) }{  \Phi_b \left (\frac{\mathsf x+\gamma}{2\pi b}+\frac{m+1}{2(m+n+1)}\ri b \right )}, \\
	F^*(\mathsf x) &= \re^{-\frac{1}{2(m+n+1)} \mathsf x} \frac{ \Phi_b \left (\frac{\mathsf x+\gamma}{2\pi b}-\frac{m+1}{2(m+n+1)}\ri b \right ) }{ \Phi_b \left (\frac{\mathsf x+\alpha}{2\pi b}+\frac{n}{2(m+n+1)}\ri b \right )  \Phi_b \left (\frac{\mathsf x+\beta}{2\pi b}+\frac{n}{2(m+n+1)}\ri b \right  ) }, \\
\ea
\ee
where $\alpha,\beta,\gamma, m,n$  are real numbers. Then, we have the following factorisation:
\be
\ba
	\label{factor5term}
	F(\mathsf x)\re^{\frac{n}{m+n+1}\mathsf y }(1+\re^{-\mathsf y})F^*(\mathsf x) &= \re^{\mathsf x'}+\re^{\mathsf y'}+\re^{\gamma} \re^{-m \mathsf x'-n \mathsf y'} \\
	&  \qquad \,\,\, +(\re^{\alpha}+\re^{\beta}) \re^{-(m+1)\mathsf x'-(n-1)\mathsf y'}+\re^{\alpha+\beta}\re^{-2(m+1) \mathsf x'-(2n-1)\mathsf y'}.
\ea
\ee
The relation between $\mathsf x, \mathsf y$ and $\mathsf x', \mathsf y'$ is 
\be
	\label{linTransf}
	\begin{pmatrix}
	\mathsf x \\ \mathsf y
	\end{pmatrix}
	=
	\begin{pmatrix}
	\mathsf -(m+1) \quad & -n \\ 1 \quad & -1
	\end{pmatrix}
	\begin{pmatrix}
	\mathsf x' \\ \mathsf y'
	\end{pmatrix},
\ee
so 
\be
	[\mathsf x',\mathsf y'] = \ri  \hbar, \qquad \qquad  \hbar= \frac{2\pi b^2}{m+n+1}.
\ee
The factorized form (\ref{factor5term}) is readily inverted:
\be
	\label{inverseFactorized}
\ba
	 \rho & \equiv
	 \left [\re^{\mathsf x'}+\re^{\mathsf y'}+\re^\gamma \re^{-m \mathsf x'-n \mathsf y'} 
	+(\re^{\alpha}+\re^{\beta}) \re^{-(m+1)\mathsf x'-(n-1)\mathsf y'}+\re^{\alpha+\beta}\re^{-2(m+1) \mathsf x'-(2n-1)\mathsf y'} \right ]^{-1}\\
	&=  \frac{1}{F^*(\mathsf x)} \left ( \frac{\re^{\frac{m+1}{m+n+1}\mathsf y }}{1+\re^{\mathsf y}} \right )\frac{1}{F(\mathsf x) }.
\ea
\ee

The three-term family of \cite{Kashaev:2015kha} is recovered in the limit $\alpha,\beta \rightarrow -\infty$ (using that  \mbox{$\Phi_b(x)  \rightarrow 1$} when \mbox{${\rm Re}(x) \rightarrow -\infty$}) and $\gamma=0$:
\be
\ba
	&\left [\re^{\mathsf x'}+\re^{\mathsf y'} +\re^{-m \mathsf x'-n \mathsf y' } \right ]^{-1} \\
	&\qquad \qquad = \re^{\frac{1}{2(m+n+1)}\mathsf x} \frac{1}{ \Phi_b \left (\frac{\mathsf x}{2\pi b}-\frac{m+1}{2(m+n+1)}\ri b \right )}
	 \left ( \frac{\re^{\frac{m+1}{m+n+1}\mathsf y }}{1+\re^{\mathsf y}} \right ) \Phi_b \left (\frac{\mathsf x}{2\pi b}+\frac{m+1}{2(m+n+1)}\ri b \right )
 \re^{\frac{1}{2(m+n+1)}\mathsf x} .
\ea
\ee
The particular case where $(m,n)=(1,1)$ gives the local $\mathbb P^2$ operator. 

The four term family of \cite{Codesido:2016ixn} is recovered when $\beta \rightarrow -\infty$ and $\gamma=0$:
\be
\ba
	&\left [\re^{\mathsf x'}+\re^{\mathsf y'} +\re^{-m \mathsf x'-n \mathsf y' }+\re^{\alpha} \re^{-(m+1)\mathsf x'-(n-1)\mathsf y'} \right ]^{-1} \\
	&\qquad \qquad = \re^{\frac{1}{2(m+n+1)}\mathsf x} \frac{ \Phi_b \left (\frac{\mathsf x+\alpha}{2\pi b}+\frac{n}{2(m+n+1)}\ri b \right ) }{ \Phi_b \left (\frac{\mathsf x}{2\pi b}-\frac{m+1}{2(m+n+1)}\ri b \right )}
	 \left ( \frac{\re^{\frac{m+1}{m+n+1}\mathsf y }}{1+\re^{\mathsf y}} \right ) \frac{\Phi_b \left (\frac{\mathsf x}{2\pi b}+\frac{m+1}{2(m+n+1)}\ri b \right )}{ \Phi_b \left (\frac{\mathsf x+\alpha}{2\pi b}-\frac{n}{2(m+n+1)}\ri b \right ) }
 \re^{\frac{1}{2(m+n+1)}\mathsf x} .
\ea
\ee
It can be seen as a perturbation of the three term operator. The particular case where $(m,n)=(0,1)$ and $\re^\alpha=m_{\mathbb F_0}$ gives the local $\mathbb F_0$ (= local $\mathbb P^1 \times \mathbb P^1$) operator studied in \cite{Kashaev:2015wia}.

A third subfamily of four term operators can be extracted from the general case. Let us set $\beta=-\alpha$ and send $\gamma \rightarrow -\infty$. We obtain
\be
\ba
	&\left [\re^{\mathsf x'}+\re^{\mathsf y'} +\re^{-2(m+1)\mathsf x'-(2n-1)\mathsf y'} + 2\cosh(\alpha) \re^{-(m+1)\mathsf x'-(n-1)\mathsf y'} \right ]^{-1} \\
	&\qquad \qquad = \re^{\frac{1}{2(m+n+1)}\mathsf x} \Phi_b \left (\frac{\mathsf x+\alpha}{2\pi b}+\frac{n}{2(m+n+1)}\ri b \right )\Phi_b \left (\frac{\mathsf x-\alpha}{2\pi b}+\frac{n}{2(m+n+1)}\ri b \right )
	 \left ( \frac{\re^{\frac{m+1}{m+n+1}\mathsf y }}{1+\re^{\mathsf y}} \right ) \\
	& \qquad \qquad  \qquad \qquad  
	 \cdot 
	\frac{1}{\Phi_b \left (\frac{ \mathsf x+\alpha}{2\pi b}-\frac{n}{2(m+n+1)}\ri b \right )
	 \Phi_b \left (\frac{\mathsf x-\alpha}{2\pi b}-\frac{n}{2(m+n+1)}\ri b \right )} \re^{\frac{1}{2(m+n+1)}\mathsf x} .
\ea
\ee
The special case with $(m,n)=(0,1)$ and $2\cosh(\alpha)=m_{\mathbb F_2}$ gives the local $\mathbb F_2$ operator. It is known that this operator is equivalent to the local $\mathbb F_0$ operator up to a unitary conjugation, a constant shift of $\mathsf x$ and an overall rescaling \cite{Kashaev:2015wia}. We see here that this is the case for the whole two families of four term operators: one of the families can be retrieved from the other by a unitary conjugation, a constant shift of  $\mathsf x$  and an overall rescaling. The unitary operator involved is given by an appropriate product of Faddeev quantum dilogarithms only depending on $\mathsf x$ (not on $\mathsf y$).

For appropriate restrictions on the values of $\alpha, \beta, \gamma$ and $m,n$, the operator $\rho$ is of trace class. This can be shown rigorously along the lines of \cite{Kashaev:2015kha}, where the proof is given for the family of three term operators and the local $\mathbb F_0$ operator.

The factorized form of our operator allows us to obtain the integral kernel of the inverse operator (\ref{inverseFactorized}). Let us introduce the basis $| x \rangle$ diagonalising the operator $\mathsf x$,
\be
	\mathsf x | x \rangle = x | x \rangle.
\ee
and the $|y \rangle$ basis diagonalising $\mathsf y$:
\be
	\mathsf y | y ) = y | y ).
\ee
We have 
\be
	\langle x | y ) = \frac{1}{2\pi b} \re^{\ri \frac{xy}{2\pi b^2 }}.
\ee
In the $x$ basis, the integral kernel of (\ref{inverseFactorized}) can be obtain for $0<\frac{m+1}{m+n+1}<1$. It is given by
\be
\ba
	\rho(x_1,x_2) &= \langle x_1 | \frac{1}{F^*(\mathsf x)} \left ( \frac{\re^{\frac{m+1}{m+n+1}\mathsf y }}{1+\re^{\mathsf y}} \right )\frac{1}{F(\mathsf x) } | x_2 \rangle  \\
	&=  \frac{1}{F^*( x_1)} \frac{1}{2\hbar  \cosh \left (\frac{ x_1-x_2}{2b^2}-\ri \pi C \right )} \frac{1}{F( x_2) },
\ea
\ee
where
\be
	C = \frac{m-n+1}{2(m+n+1)}.
\ee
It is also convenient to perform a rescaling of the our variables by defining
\be
	\nu_i = b^{-2} x_i,
\ee
as well as
\be
	\alpha = b^2 \tilde \alpha, \qquad
	\beta = b^2 \tilde \beta, \qquad
	\gamma = b^2 \tilde \gamma.
\ee
The integral kernel of $ \rho$ has to be transformed according to 
\be
	 \rho(\nu_1,\nu_2) \rd \nu_2 =  \rho(x_1,x_2) \rd x_2.
\ee
It is of the form
\be
	\label{rhoOfnu}
	 \rho(\nu_1,\nu_2) =\frac{1}{2\pi} \, \frac{\re^{-\frac{ \hbar}{2}V(\nu_1) }  \re^{-\frac{ \hbar}{2}\overline{V(\nu_2)} } }{2 \cosh \left ( \frac{\nu_1-\nu_2}{2}-\ri \pi C \right )},
\ee 
where
\be
	\label{Vfull}
\ba
	V(\nu) &=
		-\frac{1}{2\pi} \nu  + \frac{2}{ \hbar} \log \frac{ \Phi_b \left ( \frac{b}{2\pi} \left ( \nu+\tilde \alpha+\frac{\ri \pi n}{m+n+1}\right ) \right )  
		\Phi_b \left ( \frac{b}{2\pi} \left ( \nu+\tilde \beta+\frac{\ri \pi n}{m+n+1}\right ) \right )  
		 }
		  { \Phi_b \left ( \frac{b}{2\pi} \left ( \nu+\tilde \gamma-\frac{\ri \pi (m+1)}{m+n+1}\right ) \right )  
		 }.
\ea
\ee
We recall that the parameter $b$ is related to $ \hbar$ as $ \hbar = \frac{2\pi b^2}{m+n+1}$. 

We will be interested to the large $ \hbar$ (equivalently large $b$) regime of our operator. This limit is taken with fixed $\nu_i, \tilde \alpha, \tilde \beta,\tilde \gamma$.
The function $V(\nu)$ has a large $\tilde \hbar$ asymptotic expansion which can be obtained using (\ref{largebAsymp}):
\be
	\label{Vexpansion}
	V(\nu) = \sum_{k=0}^\infty  \hbar^{-2k} V_k(\nu),
\ee
with
\be
	\label{V0part}
\ba
	V_0(\nu) &= -\frac{1}{2\pi}\nu \\
	& \quad +\frac{m+n+1}{2\pi^2 \ri} 
	\left (- {\rm Li}_2(-\re^{\nu+\tilde \alpha+\frac{\ri \pi n}{m+n+1}})-
	{\rm Li}_2(-\re^{\nu+\tilde \beta+\frac{\ri \pi n}{m+n+1}})
	+ {\rm Li}_2(-\re^{\nu+\tilde \gamma-\frac{\ri \pi (m+1)}{m+n+1}}) \right ), \\[0.2cm]
\ea
\ee
and
\be
	\label{Vkpart}
\ba
	V_{k}(\nu) &= \frac{(-16 \pi^4)^k}{2\pi^2 \ri} (m+n+1)^{1-2k}\frac{(2^{-2k+1}-1)B_{2k}}{(2k)!}  \left (- {\rm Li}_{2-2k}(-\re^{\nu+\tilde \alpha+\frac{\ri \pi n}{m+n+1}})+ \right. \\
	& \qquad \qquad \qquad \qquad \qquad \qquad 
	\left.
	-{\rm Li}_{2-2k}(-\re^{\nu+\tilde \beta+\frac{\ri \pi n}{m+n+1}})
	+{\rm Li}_{2-2k}(-\re^{\nu+\tilde \gamma-\frac{\ri \pi (m+1)}{m+n+1}}) \right ) \\[0.2cm]
\ea
\ee
for $k>0$.

Now that we have an explicit expression for the integral kernel of the inverse operator $\rho$, We can explicitly obtain some relevant spectral quantities. Two important sets of quantities are the fermionic spectral traces, and the one-particle correlation functions. 
They are defined using the operator $\rho$, and, using Fredholm theory for integral operators, they can be expressed as multi-integrals involving the integral kernel $ \rho(\nu_1,\nu_2)$. Let us introduce an expansion parameter $\kappa$. The fermionc spectral traces $Z_N$ are the coefficients of the Fredholm determinant:
\be
	{\rm det}(1+\kappa   \rho ) = 1+\sum_{N=1}^\infty \kappa^N Z_N.
\ee
Using the Fredholm formula (as done for example in \cite{Marino:2015ixa}), the following expression can be obtained for $Z_N$:
\be
	\label{ZnMM}
\ba
	Z_N &=  \frac{1}{N!} \int_{\mathbb R^N} \rd^N \nu  \,\, \underset{i,j=1,...,N}{\rm det} \,\,  \rho(\nu_i,\nu_j) \\
	&= \frac{1}{N!} \int_{\mathbb R^N} \frac{\rd^N \nu}{(2\pi )^N} \re^{- \hbar \sum_{k=1}^N {\rm Re} \, V(\nu_k) } \frac{\prod_{i>j}(2\sinh \frac{\nu_i-\nu_j}{2})^2}{\prod_{i,j} 2\cosh( \frac{\nu_i-\nu_j}{2}-\ri \pi C)}.
\ea
\ee
This has the form of a partition function for a non-interacting Fermi gas with density matrix $ \rho$ (first line), or a deformed $O(2)$ matrix model written in eigenvalue variables (second line). This precise deformation of the $O(2)$ matrix model is essentially what is considered in \cite{Kostov:1999qx}. The strict $O(2)$ case is given for $C=0$. To go from the first expression to the second, we used the Cauchy determinant identity
\be
	\label{CauchyDet}
	 \underset{i,j=1,...,N}{\rm det} \,\, \frac{1}{2 \cosh \left ( \frac{\mu_i-\nu_j}{2} \right )} = \frac{\prod_{i>j} 2\sinh \frac{\mu_i-\mu_j}{2} 2\sinh \frac{\nu_i-\nu_j}{2}}{\prod_{i,j} 2 \cosh \frac{\mu_i-\nu_j}{2}}.
\ee
The matrix model form is better suited to study the large $N$ regime of $Z_N$.

Another spectral quantity which turns out to be interesting to study in our context \cite{Marino:2016rsq, Marino:2018elo} is the resolvent of (the inverse of) $ \rho$: 
\be
	\mathsf R = \frac{ \rho}{1+\kappa \rho}.
\ee
According to Fredholm theory, its integral kernel in variable $u$ has the following $\kappa$ expansion:
\be
	R(\nu,\nu') = \frac{1}{{\rm det}(1+\kappa   \rho)} \sum_{N=0}^\infty \kappa^N B_N(\nu,\nu'),
\ee
where
\be
	B_N(\nu,\nu') =  \frac{1}{N!} \int_{\mathbb R^N} \rd^N \mu \,\,  \rho 
	\begin{pmatrix}
		\nu & \mu_1 & \mu_2 & ... & \mu_N \\ 
		\nu' & \mu_1 & \mu_2 & ... & \mu_N \\ 
	\end{pmatrix}.
\ee
We used the following notation:
\be
	 \rho 
	\begin{pmatrix}
		x_1 & x_2 & ... & x_N \\ 
		y_1 & y_2 & ... & y_N \\ 
	\end{pmatrix}
	=
	\underset{i,j=1,...,N}{\rm det} \,\, \rho (x_i,y_j).
\ee
The quantities $B_N(u,u')$ have a precise interpretation in the Fermi gas picture as the unnormalized canonical reduced density matrix of a gas of $N+1$ particles. This interpretation in our context is studied in \cite{Marino:2018elo}. As for the fermionic spectral traces, the Cauchy determinant identity allows us to write $B_N$ as a correlator of a deformed $O(2)$ matrix model in eigenvalue variables.
Let us define the following expectation value at fixed $N$
\be
	\label{expVal}
	\langle f(\nu_1,...,\nu_N) \rangle = \frac{1}{Z_N}  \frac{1}{N!} \int_{\mathbb R^N} \frac{\rd^N \nu}{(2\pi )^N}  f(\nu_1,...,\nu_N) \re^{- \hbar \sum_{k=1}^N {\rm Re}\, V(\nu_k) } \frac{\prod_{i>j}(2\sinh \frac{\nu_i-\nu_j}{2})^2}{\prod_{i,j} 2\cosh \left ( \frac{\nu_i-\nu_j}{2}-\ri \pi C \right )},
\ee
and the function
\be
	t_C(\nu) = \frac{ \re^{-\ri \pi C}  \sinh(\frac{\nu}{2})}{ \cosh(\frac{\nu}{2}-\ri\pi C)}.
\ee
Then, we have that
\be
	\frac{B_N(\nu,\nu')}{Z_N} = \re^{2\pi \ri C N}  \rho(\nu,\nu') \Big  \langle \prod_{k=1}^N t_C \left ( \nu-\nu_k \right ) t_C \left (  \nu'-\nu_k \right ) \Big \rangle.
\ee
As shown in \cite{Marino:2018elo}, the 't Hooft expansion of this quantity can be written using integrals of the $n$-point connected correlators $W_{n}(X_1,...,X_n)$, as defined in eq. (\ref{nCorrW}). This is partly the reason why we are interested in the $n$-point connected correlators.

\sectiono{The gaussian expansion}  
\label{appC}
 
We often perform checks of the exact results obtained in the main body of the text against small 't Hooft coupling expansions directly obtained from the matrix model.
For completeness' sake, in this Appendix we give a rather detailed description of how we compute the small $\lambda$ expansion of these quantities. The procedure is basically perturbation theory at finite $N$ around the minimum of the potential (producing a large $\hbar$ expansion), and then after extrapolation to arbitrary $N$, a reorganization of this large $\hbar$ expansion into a t' Hooft expansion (which is possible if the quantity computed indeed admits a 't Hooft expansion). 

Let us start with an unnormalized expectation value at fixed finite $N$:
\be
\ba	
	{\Big \langle \hspace{-0.15cm} \Big  \langle}
	f(\nu_1, ... ,\nu_N)
	{\Big \rangle \hspace{-0.15cm} \Big  \rangle}
	&= \frac{1}{N!} \int_{\mathbb R^N} \frac{\rd^N \nu}{(2\pi)^N} \re^{-\hbar \sum_{k=1}^N  {\rm Re} \, V(\nu_k) } f(\nu_1,...,\nu_N) \frac{\prod_{i<j}(2\sinh \frac{\nu_i-\nu_j}{2})^2}{\prod_{i,j} 2\cosh( \frac{\nu_i-\nu_j}{2}-\ri \pi C)} \\
	&= \frac{ \re^{-N^2 \log 2\cos(\pi C)} }{N! (2\pi)^N} \int_{\mathbb R^N}  \rd^N \nu \,\,\ \re^{-\hbar \sum_{k=1}^N  {\rm Re} \, V(\nu_k) } f(\nu_1,...,\nu_N) \prod_{i<j}D(\nu_i-\nu_j),
\ea
\ee
where
\be
	D(\nu_i-\nu_j) = \frac{4\cos(\pi C)^2 \sinh^2(\frac{\nu_i-\nu_j}{2})}{\cosh(\frac{\nu_i-\nu_j}{2}+\ri \pi C)\cosh(\frac{\nu_i-\nu_j}{2}-\ri \pi C)} = (\nu_i-\nu_j)^2 \left ( 1+O(\nu_i-\nu_j)^2 \right )
\ee
is a perturbed Vandermonde determinant\footnote{Of course, the technique explained in this Appendix also works for more general perturbations of the Vandermonde determinant.}.
The potential (which has itself a large $\hbar$ expansion) can be expanded around the minimum $\nu_{\rm min}$ of its leading part:
\be
	{\rm Re} \, V(\nu) = {\rm Re} \, V_0(\nu_{\rm min}) + \frac{1}{2} {\rm Re} \, V_0''(\nu_{\rm min}) (\nu-\nu_{\min})^2 + {\sum_{g \geq 0,k \geq 0}}\hspace{-0.2cm}' \hspace{0.2cm}\frac{1}{k!} 
	{\rm Re} \, V_g^{(k)}(\nu_{\rm min}) \hbar^{-2g}(\nu-\nu_{\rm min})^k,
\ee
where the prime on the sum means that we omit the terms $(g,k)=(0,0),(0,1),(0,2)$. We perform the change of variables $ \nu_i = \tilde \nu_i \hbar^{-1/2}+\nu_{\rm min}$, and expand everything in large $\hbar$. The outcome has the form
\be
	\label{bigsumExpansion}
\ba
	{\Big \langle \hspace{-0.15cm} \Big  \langle}
	f(\nu_1, ... ,\nu_N)
	{\Big \rangle \hspace{-0.15cm} \Big  \rangle}
	& \sim
	\re^{-N^2 \log 2\cos(\pi C)-\log \Gamma(N+1)-\frac{N(N+1)}{2}\log\hbar -N \log 2\pi-\hbar N \,{\rm Re}\, V_0(\nu_{\min})}\\
	 & 
	 \qquad \qquad
	 \times  \sum_{k \geq 0} \hbar^{-k/2} \sum_{\boldsymbol \ell} c_{k,{\boldsymbol \ell}}(N)
	 \prod_{j=1}^N \int_{\mathbb R} \rd {\tilde \nu}_j \re^{-\frac{1}{2} {\rm Re} \, V_0''(\nu_{\rm min})} {\tilde \nu}_j^{\ell_j}.
\ea
\ee
where $\boldsymbol \ell=(\ell_1,...,\ell_N)$ is a vector of positive integers, and the coefficients $c_{k,{\boldsymbol \ell}}(N)$ are obtained by putting together the large $\hbar$ expansions of the different pieces. All the one dimensional gaussian integrals are readily performed using that for $a>0$ and $\ell$ a positive integer,
\be
	\int_{\mathbb R} \rd \nu \, \re^{-\frac{1}{2}a\nu^2} \, \nu^\ell = \frac{1+(-1)^\ell}{2}\Gamma \left (\frac{\ell+1}{2} \right )\left ( \frac{2}{a} \right )^{\frac{\ell+1}{2}}.
\ee
When $k$ is odd in (\ref{bigsumExpansion}), it can be seen that some of the $\ell_j$ are odd integers, so those terms vanish. The result takes therefore the form
\vspace{0.2cm}
\be
\ba
	{\Big \langle \hspace{-0.15cm} \Big  \langle}
	f(\nu_1, ... ,\nu_N)
	{\Big \rangle \hspace{-0.15cm} \Big  \rangle}
	& \sim
	\re^{-N^2 \log 2\cos(\pi C)-\log \Gamma(N+1)-\frac{N(N+1)}{2}\log\hbar-\hbar N \,{\rm Re}\, V_0(\nu_{\min})-\frac{N}{2} \log \left (2\pi^2 {\rm Re}\, V_0''(\nu_{\rm min})\right )} \\
	& \qquad \qquad \qquad
	\times  \sum_{k \geq 0} \hbar^{-k} \tilde C_{k}(N).
\ea
\ee

For $f(\nu_1,...,\nu_n)=1$, we are computing the partition function. Taking the logarithm gives the free-energy. Since we expect that it admits a 't Hooft expansion, we can reorganise the large $\hbar$ series accordingly. The exponentiated prefactor is part of the gaussian contribution
\be
\ba
	 Z_N^{\rm gaussian} &=\frac{ \re^{-N^2 \log 2\cos(\pi C)-\hbar N \,{\rm Re}\, V_0(\nu_{\min})} }{\hbar^{\frac{N(N+1)}{2} }(2\pi)^N N!} \int_{\mathbb R^N} \rd^N \tilde \nu \,\, \re^{-\frac{1}{2} {\rm Re}\, V_0''(\nu_{\min}) \sum_{j=1}^N \tilde \nu_j^2 }  \prod_{i<j}(\tilde \nu_i-\tilde \nu_j)^2 \\
	 				&=\frac{ \re^{-\frac{N^2}{2} \log\left (  4\cos^2(\pi C) {\rm Re}\, V_0''(\nu_{\rm min}) \right )}}{\hbar^{\frac{N(N+1)}{2} }(2\pi)^{N/2}}G(N+1)  
\ea
\ee
which we factor out. In the above, $G(N)$ is the Barnes function. Its logarithm has a well known large $N$ expansion.
At each order in large $\hbar$, the coefficient of $\log Z_N-\log Z_N^{\rm gaussian}$ has to have a strict polynomial dependance in $N$ otherwise it cannot be converted into a 't Hooft expansion:
\be
\ba
	\log Z_N -\log Z_N^{\rm gaussian} 
	&= \log {\Big \langle \hspace{-0.15cm} \Big  \langle}
	1
	{\Big \rangle \hspace{-0.15cm} \Big  \rangle}
	-\log Z_N^{\rm gaussian} \\
	& \sim 
	\sum_{k \geq 1} \hbar^{-k}  C_{k}(N)
	 \\
	& \sim \frac{C_{1,3}N^3+C_{1,1}N}{\hbar}+\frac{C_{2,4}N^4+C_{2,2}N^2}{\hbar^2}+\frac{C_{3,5}N^5+C_{3,3}N^3+C_{3,1}N}{\hbar^3}+...
\ea
\ee
By fitting the finite $N$ values of $C_{k}(N)$ to the expected polynomial behaviour, we find the values of $C_{k,g}$. After using $N=\lambda \hbar$, plugging in the large $N$ asymptotic series of $Z_N^{\rm gaussian}$, and reorganizing the series, we obtain the 't Hooft expansion at small coupling $\lambda$:
\be
\ba
	\log Z_N   &\sim  \log Z_N^{\rm gaussian}  + \hbar^{2}\left ( C_{1,3}\lambda^3 + C_{2,4}\lambda^4 +  C_{3,5}\lambda^5+... \right ) +(C_{1,1}\lambda +C_{2,2} \lambda^2+C_{3,3}\lambda^3+...)\\
	& 
	\qquad  \qquad \qquad \,\, \,
	+\hbar^{-2} (C_{3,1}\lambda+...)+...\\
		&\sim \sum_{g=0}^\infty \hbar^{2-2g} \mathcal F_g(\lambda),
\ea
\ee
where
\be
\ba
	 \log Z_N^{\rm gaussian} &\sim \hbar^2 \left ( \frac{\lambda^2}{2} \log \frac{\lambda}{4\cos^2(\pi C) \, {\rm Re}\, V_0''(\nu_{\rm min})}-\frac{3\lambda^2}{4} \right )+\left ( -\frac{1}{12}\log \lambda +\zeta'(-1) \right ) 
	\\
	 & \qquad \qquad
	+\sum_{g=2}^\infty \hbar^{2-2g} \frac{B_{2g}}{2g(2g-1)} \lambda^{2-2g} -\frac{1}{12}\log \hbar.
\ea
\ee
We usually drop the $-\frac{1}{12}\log \hbar$ term in order to obtain a proper 't Hooft expansion.

For other functions $f(\nu_1,...,\nu_n)$ such as multi-traces or its generating functions (the $n$-point correlators), a 't Hooft expansion can be expected to be found for its normalized expectation value. We have
\be
\ba
	\Big \langle 	f(\nu_1, ... ,\nu_N) \Big \rangle
	&=
	\frac{1}{Z_N}{\Big \langle \hspace{-0.15cm} \Big  \langle}
	f(\nu_1, ... ,\nu_N)
	{\Big \rangle \hspace{-0.15cm} \Big  \rangle} \\
	&\sim \sum_{k \geq 0} \hbar^{-k}  C_{k}(N).
\ea
\ee
Again, if a 't Hooft expansion exists, the $C_{k}(N)$ are polynomials in $N$ which can be fixed from finite $N$ results. Setting $N=\lambda \hbar$ and rearranging the large $\hbar$ expansion, we obtain the 't Hooft expansion at small coupling $\lambda$. The connected correlators are also easily computed in this way.

 \sectiono{A formula for  the function $\lambda(q)$}
 \label{appD}
 
In this Appendix, we show how expression (\ref{lambdaGeneral}) for $\lambda(\tau)$ can be simplified to expression (\ref{lambdaNice2}). The second formula is quite interesting, since it only involves $u_\alpha,u_\beta,u_\gamma$ (corresponding to  the singularities of the potential) and the conformal map $X(u)$. We start by noticing that (\ref{lambdaGeneral})  is basically made of three distinct parts, which we may write as
 \be
 	\lambda = \frac{m+n+1}{8\pi^3} \left[ L(u_\gamma)-L(u_\alpha)-L(u_\beta) \right ],
 \ee
 where
 \be
 	L(u_i) = \int_{-\frac{1}{2}-\frac{\tau}{2} }^{\frac{1}{2}-\frac{\tau}{2}} \rd u \left (  \frac{\vartheta_1'}{\vartheta_1}(u_\infty+u) +  \frac{\vartheta_1'}{\vartheta_1}(u_\infty-u) \right ) 
	\left ( \log  \frac{\vartheta_1(u-u_i)}{ \vartheta_1(u_\infty+u_i)}-\log \frac{\vartheta_1(u-u_\infty)}{ \vartheta_1(2u_\infty) }  \right ).
 \ee
 We will express this integral in another way. Let us start by differentiating $L(u_i)$ with respect to $u_i$. We find after some simplifications:
 \be
 \ba
 	L'(u_i) &= - \int_{-\frac{1}{2}-\frac{\tau}{2} }^{\frac{1}{2}-\frac{\tau}{2}} \left (  \frac{\vartheta_1'}{\vartheta_1}(u_\infty+u) +  \frac{\vartheta_1'}{\vartheta_1}(u_\infty-u) \right )  \frac{\vartheta_1'}{\vartheta_1}(u-u_i), \\
	L''(u_i) &= \int_{-\frac{1}{2}-\frac{\tau}{2} }^{\frac{1}{2}-\frac{\tau}{2}} \left (  \frac{\vartheta_1'}{\vartheta_1}(u_\infty+u) +  \frac{\vartheta_1'}{\vartheta_1}(u_\infty-u) \right ) \left ( \frac{\vartheta_1'}{\vartheta_1} \right )'(u-u_i)
\ea
 \ee
In the second integral, the integrand consists of a product of two elliptic functions (the first is elliptic as we already argued in the main text, and the second is the Weierstrass $\wp$ function up to a constant). Their product is therefore an elliptic function of $u$ which can be obtained by looking at its polar behaviour. We need an elliptic function with poles of degree $1$ at $u=\pm u_\infty$ and poles of degree up to two at $u=u_i$. The function is found to be
\be
	\label{prodEllipticDDL}
\ba
	\left (  \frac{\vartheta_1'}{\vartheta_1}(u_\infty+u) +  \frac{\vartheta_1'}{\vartheta_1}(u_\infty-u) \right ) \left ( \frac{\vartheta_1'}{\vartheta_1} \right )'(u-u_i) &= \alpha_1 \frac{\vartheta_1'}{\vartheta_1}(u_\infty+u) +\alpha_2 \frac{\vartheta_1'}{\vartheta_1}(u_\infty-u)  \\
	& \qquad
	+(\alpha_2-\alpha_1)  \frac{\vartheta_1'}{\vartheta_1}(u-u_\gamma) \\
	& \qquad +\alpha_3 \left ( \frac{\vartheta_1'}{\vartheta_1} \right )'(u-u_i) +C,
\ea
\ee
where $\alpha_i$ are obtained comparing the polar behaviours of the lefthand and righthand sides,
\be
\ba
	\alpha_1 &= \left ( \frac{\vartheta_1'}{\vartheta_1} \right )'(u_\infty+u_i), 
	\qquad
	 \alpha_2 = \left ( \frac{\vartheta_1'}{\vartheta_1} \right )'(u_\infty-u_i), \\
	 \alpha_3 &= \left (  \frac{\vartheta_1'}{\vartheta_1}(u_\infty+u_i) +  \frac{\vartheta_1'}{\vartheta_1}(u_\infty-u_i) \right ),
\ea
\ee
and the constant $C$ is obtained by evaluating the equality at any point $u$. We do it for $u=u_i$, and obtain the corresponding expression for $C$ (the poles cancel and we equate the finite parts). We find that it can be compactly written as
\be
	C=\frac{1}{2} \frac{\rd}{\rd u_i} \left ( \frac{\vartheta_1''}{\vartheta_1}(u_\infty-u_i)-\frac{\vartheta_1''}{\vartheta_1}(u_\infty+u_i)  \right ).
\ee
Using the expression (\ref{prodEllipticDDL}), the integral over $u$ for $L''(u_i)$ can be easily obtained. Actually, none of the functions contribute (their integral vanish or compensate along that path): only the constant $C$ gives a non vanishing contribution. We end up with 
\be
	L''(u_i) = C.
\ee
 We can integrate this up twice and use that $L(u_\infty)=0$ (directly from the definition of $L(u_i)$) and  $L'(0)=0$ (from symmetry considerations). Therefore, we get a new integral expression for $L(u_i)$
 \be
 	L(u_i) = \frac{1}{2} \int_{u_\infty}^{u_i} \rd u \left (  \frac{\vartheta_1''}{\vartheta_1}(u_\infty-u)-\frac{\vartheta_1''}{\vartheta_1}(u_\infty+u) \right ).
 \ee
 As all $\vartheta$-functions, $\vartheta_1(u,\tau)$ as a function of the two variables $u,\tau$ obeys the well known diffusion equation:
 \be
 	\vartheta_1''(u,\tau) = 4\pi \ri \frac{\partial}{\partial \tau} \vartheta_1(u,\tau),
 \ee
 where primes denote differentiation with respect to $u$. We therefore obtain the neat relation
 \be
 	L(u_i) = -2\pi \ri \int_{u_\infty}^{u_i} \rd u \frac{\partial}{\partial \tau} \log \frac{\vartheta_1(u_\infty+u,\tau)}{\vartheta_1(u_\infty-u,\tau)} =  -2\pi \ri \int_{u_\infty}^{u_i} \rd u \frac{\partial}{\partial \tau} \log  X(u).
 \ee 
 This implies the following remarkable relation for $\lambda$:
 \be
 	\label{lambdaNice}
 	\lambda = \frac{m+n+1}{4\pi^2 \ri} \left (  \int_{u_\infty}^{u_\gamma}-  \int_{u_\infty}^{u_\alpha}-  \int_{u_\infty}^{u_\beta} \right )\rd u  \frac{\partial}{\partial \tau} \log  X(u).
 \ee
 The differentiation should only be applied on the second variable of theta functions $\vartheta_1(u,\tau)$ which compose the map $X(u)$ (this means for example that here we consider $\sqrt{ab}$ to be independent of $\tau$ when differentiating).
The locations of our potential's singularities $u_{\alpha,\beta,\gamma}$ in the $u$ plane are the only necessary parameters to determine the function $\lambda(\tau)$, together with the universal map $X(u)$ and the constant $u_\infty$, both of which are potential independent. We note that $u_{\alpha,\beta,\gamma}$ may have their own $\tau$ dependence, but they are not differentiated.
 
 This formula is also useful in the strong 't Hooft coupling expansion (which involves an $S$-transformation of the elliptic functions), whereas (\ref{lambdaGeneral}) is less useful in that regime because the integral does not seem to be consistent with the dual expansion.

 \sectiono{The expression for $P_{ij}(x)$}
 \label{appE}
 
 The polynomials $P_{ij}(x)$ for $i,j=0,1,...,6$ in (\ref{finalvarpi20}) can be put into a $7 \times 7$ symmetric matrix $P(x)$ whose entries are $P_{ij}(x)$:
 \footnotesize
\be
\ba
P(x) &=
\left(
\begin{array}{cccc}
 10-4 x^2 & 16 x^3+8 x & 80 x^2-50 & 32 x^3-32 x  \\
 16 x^3+8 x & 160 x^4-64 x^2 & 64 x^5+240 x^3-184 x & 0 \\
 80 x^2-50 & \quad  64 x^5+240 x^3-184 x \quad & -64 x^6+448 x^4-196 x^2-38 & \quad -128 x^5-32 x^3+160 x \quad \\
 32 x^3-32 x & 0 & -128 x^5-32 x^3+160 x & 512 x^2-512 x^4 \\
 16 x^4-76 x^2+30 & -64 x^5-240 x^3+184 x & -544 x^4+256 x^2+138 & -128 x^5-32 x^3+160 x \\
 -16 x^3-8 x & 64 x^2-160 x^4 & -64 x^5-240 x^3+184 x & 0 \\
 -6 & -16 x^3-8 x & 16 x^4-76 x^2+30 & 32 x^3-32 x \\
\end{array}
\right.
\cdots \\
& \qquad \qquad 
\cdots
\left.
\begin{array}{ccc}
 16 x^4-76 x^2+30 & -16 x^3-8 x & -6 \\
 -64 x^5-240 x^3+184 x & 64 x^2-160 x^4 & -16 x^3-8 x \\
 -544 x^4+256 x^2+138 & -64 x^5-240 x^3+184 x & \quad 16 x^4-76 x^2+30 \quad \\
 -128 x^5-32 x^3+160 x & 0 & 32 x^3-32 x \\
 \quad -64 x^6+448 x^4-196 x^2-38 \quad & 64 x^5+240 x^3-184 x & 80 x^2-50 \\
 64 x^5+240 x^3-184 x & 160 x^4-64 x^2 & 16 x^3+8 x \\
 80 x^2-50 & 16 x^3+8 x & 10-4 x^2 \\
\end{array}
\right).
\ea
\ee
 
\bibliographystyle{JHEP}
\bibliography{/Users/sz/Dropbox/UNI/BibTex/bibSZ}

\end{document}